\documentclass[aps,prb,longbibliography,twocolumn,showpacs,floatfix,superscriptaddress]{revtex4-1}

\usepackage{graphicx,color}
\usepackage{amsfonts}
\usepackage[figuresright]{rotating}  
\usepackage{amssymb}
\usepackage{amsmath}
\usepackage{mathtools}
\usepackage{psfrag}
\usepackage{subfigure}
\usepackage{multirow}
\usepackage{tabularx}
\usepackage{textcomp}
\usepackage{units}
\usepackage{lipsum}
\usepackage{soul}

\def\beq{\begin{eqnarray}}
\def\eeq{\end{eqnarray}}

\renewcommand{\v}[1]{\ensuremath{\mathbf{#1}}} 
 
\let\baraccent=\= 
\renewcommand{\=}[1]{\stackrel{#1}{=}} 
	
\makeatletter

\begin{document}
\title{Design principles for shift current photovoltaics}

\author{Ashley\ M. \surname{Cook}}
\thanks{These two authors contributed equally.}
\affiliation{Department of Physics, University of California, Berkeley,
California, 94720, USA} 
\affiliation{Department of Physics, University of Toronto, CAN}

\author{Benjamin\ \surname{M. Fregoso}}
\thanks{These two authors contributed equally.}
\affiliation{Department of Physics, University of California, Berkeley,
California, 94720, USA}

\author{Fernando\ \surname{de Juan}}
\affiliation{Department of Physics, University of California, Berkeley,
California, 94720, USA}

\author{Sinisa\ \surname{Coh}}
\thanks{Current address: Mechanical Engineering, Materials Science and Engineering, University of California Riverside, Riverside, CA 92521, USA}
\affiliation{Department of Physics, University of California, Berkeley,
California, 94720, USA}

\author{Joel\ E.  \surname{Moore}}
\affiliation{Department of Physics, University of California, Berkeley,
California, 94720, USA}
\affiliation{Materials Sciences Division,
Lawrence Berkeley National Laboratory, Berkeley, CA 94720}


\begin{abstract}
While the basic principles of conventional solar cells are well understood, little attention has gone toward maximizing the efficiency of photovoltaic devices based on shift currents. By analyzing effective models, here we outline simple design principles for the optimization of shift currents for frequencies near the band gap. Our method allows us to express the band edge shift current in terms of a few model parameters and to show it depends explicitly on wavefunctions in addition to standard band structure. We use our approach to identify two classes of shift current photovoltaics, ferroelectric polymer films and single-layer orthorhombic monochalcogenides such as GeS, which display the largest band edge responsivities reported so far. Moreover, exploring the parameter space of the tight binding models that describe them we find photoresponsivities that can exceed 100 mA W$^{-1}$. Our results illustrate the great potential of shift current photovoltaics to compete with conventional solar cells.
\end{abstract}



\maketitle
\textit{\textbf{Introduction}} - Cost-effective, high-performing solar cell technology is an essential piece 
of a sustainable energy strategy.  Exploring approaches to photo-current generation beyond conventional 
solar cells based on pn junctions is worthwhile given that their performance is in practice 
constrained by the Shockley-Queisser limit\cite{Shockley1961}.  One of the most promising alternative 
sources of photocurrent is the bulk photovoltaic effect (BPVE) or `shift current' effect, a non-linear 
optical response that yields net photocurrent in materials with net polarization~\cite{KB79,BS80,Baltz1981,PB82,Sturman1992,AS95,KBH82,Sipe2000,KMT00}: Contrary to conventional pn junctions, the BPVE is able to generate an above band-gap 
photovoltage~\cite{bpve_abovebgvoltage}, potentially allowing the performance of BPVE-based photovoltaics 
to surpass conventional ones. However, closed-circuit currents generated via the BPVE reported in the literature have typically been small compared to those generated in pn junction photovoltaics\cite{ZTW15,YR12,BYZ14}. Recent interest in the BPVE also stems from the proposal that it may be at 
work in a promising class of materials for photovoltaics known as hybrid perovskites~\cite{ZTW15}, an 
extremely active field of research~\cite{Hodes2013,Egger2015, hybperov_improvement, hybperov_cheap, hybperov_soc001, hybperov_soc002, hybperov_soc003, hybperov_excitoncorr, hybperov_dis, hybperov_abinit001, hybperov_abinit002, hybperov_ex001,hybperov_ex002, hybperov_ex003}.

The fundamental requirement for a material to produce a current via the BPVE is that it breaks inversion symmetry, allowing an asymmetric photoexcitation of carriers. But despite considerable case-by-case study of the BPVE, the necessary ingredients to optimize a BPVE-based solar cell are not sufficiently well understood. As with conventional solar cells, band gaps in the visible (1.1-3.1 eV) ~\cite{YZR12, BYZ14} and large electronic densities of states~\cite{YR12, rappe_layeredferro} are always beneficial. In addition, to produce a solar cell that responds to unpolarized sunlight, a highly anisotropic material must be used, since otherwise there is no preferred direction for the current to flow. But beyond these natural requirements, our only guiding knowledge is that the shift current depends explicitly on the nature of the electronic wavefunctions~\cite{Wang2015, rappe_layeredferro} and that it is not correlated with the material polarization in any obvious way~\cite{BYZ14} despite the fact that both shift currents and polarization originate from inversion symmetry breaking.

\begin{figure}[t!]
\includegraphics[width=8.5cm]{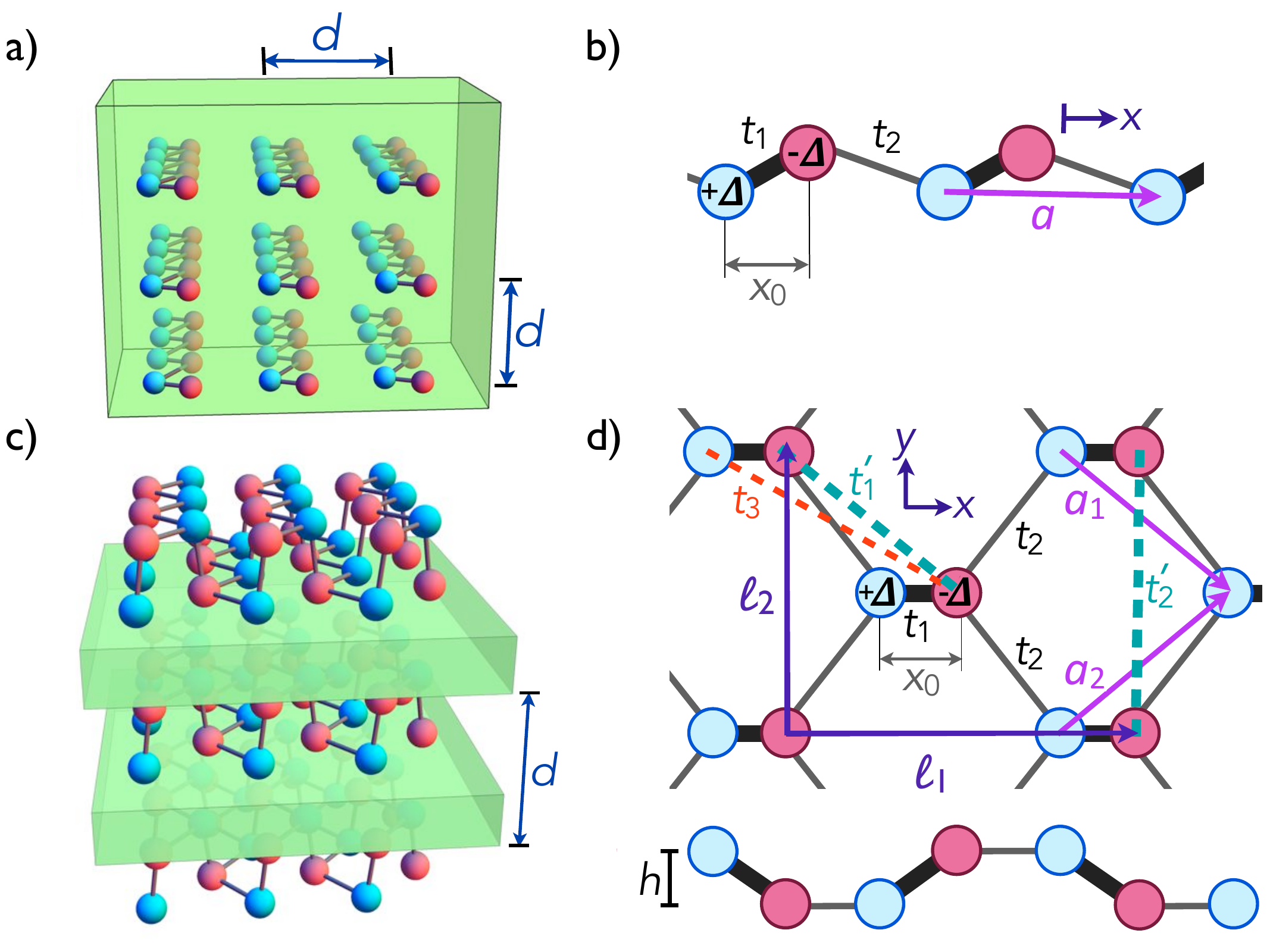}
\caption{Schematics of proposed shift current photovoltaics: a) 3D structure of a solar cell built by stacking one-dimensional ferroelectric polymers. 
b) Simplified two-band tight binding model of a polymer. c) 3D structure of a solar cell made by stacking two-dimensional monolayers of a monochalcogenide. The inert spacers between layers prevent the restoration of bulk inversion symmetry. d) Simplified two-band tight-binding model for a monochalcogenide layer.}
\label{fig1}
\end{figure}

In the current situation, a more generic understanding of what makes the BPVE strong is highly desirable. When tackling complex material science problems, stripping off all complications and optimizing the simplest model that captures the relevant physics often proves the best strategy, as shown for example in thermoelectricity studies~\cite{mahan,disalvo,murphy_moore}. In this work, we present simple design principles for BPVE optimization based on the study of an effective model for the band edges. With this model, the band edge shift current is given by the product of the joint density of states (JDOS) and a matrix element, both given by simple expressions in terms of a few model parameters. The simplicity of the model allows us to derive the main principle that band edges with semi-Dirac type of Hamiltonians are the best starting point to obtain large band edge prefactors. In addition, by relating the effective model parameters to realistic tight-binding models, we can predict that several materials with the required band structure have larger shift currents than any reported so far.

\textit{\textbf{Results}} - In our search for materials we should look for large JDOS in systems where the band edge is closely aligned with the peak of the solar spectrum, around 1.5 eV. Since the band edge always induces a Van Hove singularity in the density of states, the requirement of a large peak in the photoresponse can be naturally better satisfied by low-dimensional materials, which generically present stronger singularities~\cite{VH53}.  Materials of one and two dimensions are therefore the focus of this work. Among one-dimensional materials, ferroelectric polymers are suitable candidates for shift-current photovoltaics: they strongly break inversion symmetry, some have suitable band gaps for photovoltaics applications\cite{Nalwa1995,L83,GFL99,RM82}, and they can be produced in macroscopically oriented samples.  For these reasons, we consider solar cells consisting of such polymer films, shown in Fig. \ref{fig1}(a). Two-dimensional materials~\cite{GG13} also have great potential for photovoltaics, as shown by demonstration of a pn-junction photovoltaic effect in dichalcogenide heterostructures \cite{BRE13,YLZ13,BPG13}, and in few-layer black phosphorus~\cite{BGS14}. However, these well known 2D semiconductors have vanishing shift currents because of either inversion or rotation symmetry. Group IV monochalcogenides have emerged in the past years as a new familiy of inversion-breaking, anisotropic 2D materials with fascinating properties~\cite{SH14,GC15,ABB11,LLW16,RGC16}, and interest is growing as thin films of all four members of the family, GeS\cite{LHS12,ULK16,VPH10,RKD16}, GeSe\cite{VPH10,RKD16}, SnS\cite{BLL15,XLH16} and SnSe\cite{LCH13,ZZW15,ZWZ15}, have now been isolated experimentally. In this work, we show that GeS is ideally suited to realize high values of the BPVE. Their GeS structure is shown in Fig.\ref{fig1}(c).

To understand how to optimize the photoresponse, we first discuss how the shift current can be computed for a tight binding model, and then we proceed to apply this formalism to describe a generic band edge and the response of particular materials.

\textbf{Shift current} - In this work we consider the shift current contribution to the BPVE and we shall use both terms interchangeably (note the BPVE can have other contributions as well\cite{Sturman1992}). With electric field $E_b(\omega)$ at frequency $\omega$ and linearly-polarized in the $b$ direction, the shift current is a DC response of the form \cite{Sturman1992}
\begin{align}
J_a = \sigma^{abb}(\omega) E_b(\omega) E_b(-\omega).
\label{shiftcurrent}
\end{align}
Defining an intensity for each polarization, $I_{0,b} = c \epsilon_0 |E_b|^2/2$, we define the photoresponsivity $\kappa^{abb}$ as the current density generated per incident intensity $J_a = \kappa^{abb} I_{0,b}$, which gives $\kappa^{abb} = 2 \sigma^{abb}/c \epsilon_0$. Note that in conventional solar cells the current is also linear with intensity. For a D-dimensional system, $\kappa^{abb}$ takes the form~\cite{AS95,Sipe2000}
\begin{align}
\kappa^{abb} =& C \int \frac{dk^D}{(2\pi)^D} \sum_{n,m} {\rm f}_{nm} I^{abb}_{nm}\delta(\omega_{nm}-\omega),
\label{kappa_abb}
\end{align}
where $C=4 g_s \pi e^3/ \hbar^2 \epsilon_0 c$, with $c$ the speed of light, $\epsilon_0$ the vacuum permittivity, 
and $g_s=2$ accounts for the spin degeneracy. In what follows we set $\hbar=1$. Summation of indices is explicitly indicated using the summation symbol. The sum is over all Bloch bands, with $\omega_{nm} = E_n-E_m$ the energy difference between bands $n$ and $m$ and ${\rm f}_{nm}={\rm f}_n-{\rm f}_m$ the difference of Fermi occupations, which we take at zero temperature. The integrand is
\begin{align}
I^{abb}_{nm} = {\rm Im}(r^b_{mn}r^b_{nm;a})
\label{integrandd},
\end{align}
where $r_{nm}^a$ are the inter-band matrix elements of the position operator (or inter-band Berry connections), defined as $r_{nm}^a = i \left<n|\partial_{k_a}m\right>$ for $n\neq m$ and zero otherwise, where $\left|n\right>$ is the eigenstate of band $n$. 
A semicolon denotes a generalized derivative $r^b_{nm;a} = \partial_{k_a}r_{nm}^b -i(\xi^a_{nn}-\xi^a_{mm})r_{nm}^b$, where $\xi_{nn}^a =i \left<n|\partial_{k_a}n\right>$ is the diagonal Berry connection for band $n$. 

\begin{figure*}[t]
\begin{center}
\includegraphics[width=18cm]{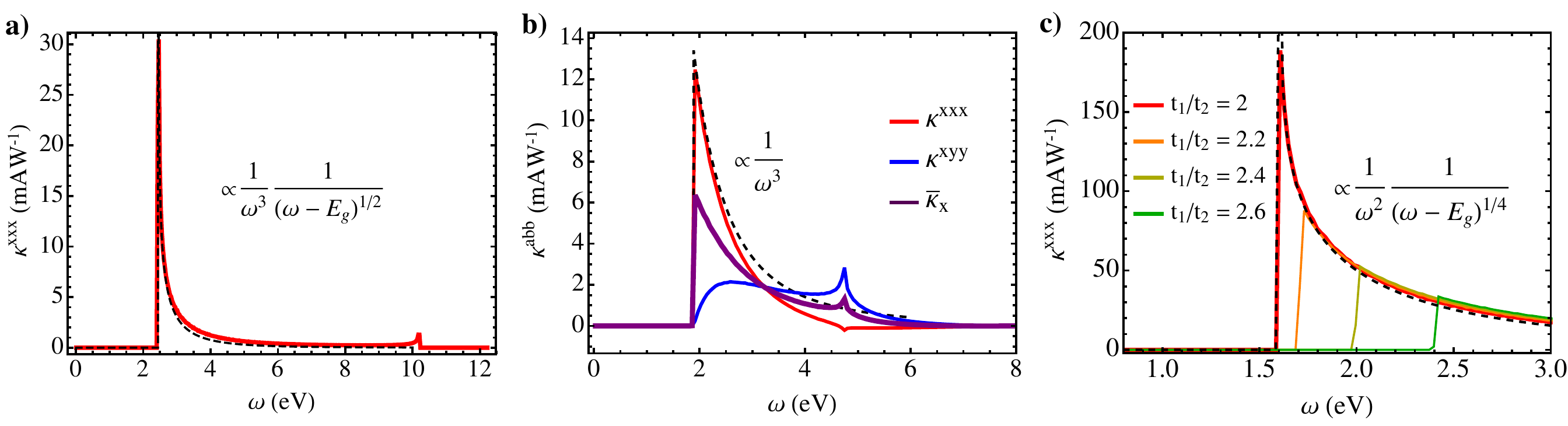}
\caption{Frequency dependence of the components of photoresponsivity $\kappa^{abb}$ for different tight-binding models, computed from Eqs.~\ref{kappa_abb} and~\ref{intsum}: (a) Responsivity for a stack of disubstituted polyacetylene polymers with tight binding parameters $t_1 = 2.85$, $t_2= 2.15$, $\Delta = 1.0$ in eV, showing the square root divergence of the current at the band edges. (b) Various non-zero components of the responsivity tensor for a stack of 2D monochacogenides with parameters $t_1 = - 2.33$, $t_2 = 0.61$, $t_3 = 0.13$, $\Delta = 0.41$ in eV, and $x_0 = 0.52 {\rm \AA}$. A large peak is observed in $\kappa^{xxx}$ at the band edge. (c) Responsivity for $\Delta = 0.8$ eV, $x_0 =0.6 {\rm \AA}$, $t_3=0$ and different hopping ratios $|t_1|/t_2$ approaching the semi-Dirac limit. The emergence of a singularity is observed. In the three figures, solid lines show the shift current components as computed from the tight-binding model, and a dashed line in each subfigure shows the $xxx$ shift current component as predicted by the effective low energy model valid near the edge, Eq.~\ref{sigma}}\label{fig:SC_vs_omega}.
\end{center}
\end{figure*}
 
\textbf{Generic two band model} - With the aim of describing the shift current response of the band edge of a semiconductor, next we consider the shift current of a generic two band model. The Fourier transform of the real space Hamiltonian is performed with the choice of phases $\psi_{\bf k}(x) = \tfrac{1}{N} \sum_{\bf R,i} e^{i \bf{k} ({\bf R}+{\bf x}_{i})}\phi({\bf x} -{\bf R}-{\bf x}_{i}) \left|m\right>_{{\bf k},i}$, where $\phi(\bf x)$ is a localized orbital and ${\bf x}_i$ is the position of site $i$ in the unit cell. This choice is made in order to naturally incorporate the action of the position operator, see Refs.~\cite{BM09,DDM15,FCG14}. The Hamiltonian matrix takes the form
\begin{equation}
H = \epsilon_0\sigma_0  + \sum_i \sigma_i f_i,
\label{eqn:gen_two_band_Ham}
\end{equation}
where $\sigma_0$ is the identity matrix, $\sigma_i=\sigma_x,\sigma_y,\sigma_z$ 
are the Pauli matrices and $\epsilon_0$ and $ f_i=f_x,f_y,f_z$ 
are generic functions of momenta $\v{k}$ (the momentum label is omitted to simplify notation). 
The conduction and valence bands are given by $E_1 = \epsilon_0 + \epsilon$, $E_2 = \epsilon_0 - \epsilon$, 
respectively and $ \epsilon = (\sum_i f_i f_i)^{1/2}$. Note that this basis choice implies that the Hamiltonian matrix elements are not periodic in the Brillouin Zone, $H_{ij}({\bf k}+{\bf G}) \neq H_{ij}({\bf k})$ with ${\bf G}$ a reciprocal lattice vector.

To compute the shift current, the direct use of Eq.~\ref{integrandd} requires the evaluation of derivatives of 
Bloch functions, which can be difficult to compute numerically. Previous works~\cite{Baltz1981,AS95,Sipe2000} 
have addressed this problem with the use of identities that replace wavefunction derivatives with sums over all 
states of matrix elements of Hamiltonian derivatives. These identities are known as sum rules and rely on the fact 
that momentum and velocity operators are proportional in the plane wave basis $p=mv$, which is not true in 
the tight binding formalism. In this work we derived a generalized sum rule appropriate for tight binding models 
(see Methods section), from which the integrand Eq.~\ref{integrandd} can be evaluated for any two-band model 
in terms of the Hamiltonian derivatives only. The result is 
\begin{equation}
I^{abb}_{12}
= -\sum_{ijm}\frac{1}{4\epsilon^3} (f_m f_{i,b} f_{j,ab}-f_m f_{i,b} f_{j,a} \frac{\epsilon_{,b}}{\epsilon})\varepsilon_{ijm},
\label{intsum}
\end{equation}
where the compact derivative notation $f_{i,a} \equiv \partial_{k_a} f_i$ and 
$\epsilon_{,b} \equiv \partial_{k_b} \epsilon$ is used. Eq. \ref{intsum} is one of the main results of this work. 
Several general principles to maximize the band edge shift current can be derived from this expression. 
A straightforward one is that, since this expression does not depend on $\epsilon_0$, particle-hole asymetry 
does not influence the shift current at all. Therefore $\epsilon_0$ is set to zero from now on. The additional term that appears only for tight binding models in this more general sum rule is $f_m f_{i,b} f_{j,ab}$, which is absent in previous formulations. For a direct band gap, this term dominates the response exactly at the band edge, since to lowest order in $k$ the first term always has constant contribution, while the second one is at least linear in $k$ for any model due to the energy derivative $\epsilon_{,b}$. For this
 term to be finite, the three Pauli matrices in the Hamiltonian must have constant, linear, and quadratic 
coefficients, in any order. Satisfying this low-energy constraint can be taken as another general principle 
in the search for materials with large shift current.

More explicit guidelines can be obtained by considering an explicit low-energy model with a direct 
band gap at a time reversal invariant momentum. Expanding the Hamiltonian around it we get
\begin{align}
H =& (\delta + \alpha_x k_x^2 + \alpha_y k_y^2+ \alpha_{xy}k_x k_y)\sigma_x \nonumber \\+& v_F k_x \sigma_y + (\Delta + \beta_x k_x^2 + \beta_y k_y^2+\beta_{xy}k_x k_y)\sigma_z.
\label{H2D}
\end{align}
Time reversal symmetry $H^*(-\v{k}) = H(\v{k})$ prevents 
quadratic terms in $\sigma_y$, and we have taken the linear term to be in the $x$ direction without loss 
of generality. Note this type of linear term requires the breaking of any $C_n$ rotation symmetry with 
$n>2$. The band gap of this model is $E_{\rm g}=2 \epsilon_{\v{k}=0}$. Evaluating \ref{intsum} we get

\begin{align}
I^{xxx}_{12}(\omega) &= \frac{4v_F}{\omega^3}(\alpha_x \Delta-\beta_x \delta) + O(\v{k}^2),\label{eqn:integrandd_close_band_edge1}\\
I^{xyy}_{12}(\omega) &= \frac{2v_F}{\omega^3} (\alpha_{xy} \Delta - \beta_{xy}\delta) + O(\v{k}^2),
\label{eqn:integrandd_close_band_edge2}
\end{align}
while $I^{yxx}_{12} = I^{yyy}_{12} = 0 + O(k^2)$. Also note that in order to have a non-zero shift current 
quadratic terms in $\sigma_x$ or $\sigma_z$ are required. In 2D, the fact that $I^{xyy}$ is in general 
non-zero means that the current need not be in the direction of the electric field polarization.

The shift current close to the band edge can now be obtained by substituting 
Eqs.~\ref{eqn:integrandd_close_band_edge1}-\ref{eqn:integrandd_close_band_edge2} into Eq.~\ref{kappa_abb}, which gives

\begin{align}
\kappa^{abb}(\omega) = C \;   I^{abb}_{12}(\omega) N(\omega), & & (\omega-E_{\rm g})/E_{\rm g} \ll 1 
 \label{sigma}
\end{align}
where $N(\omega)=\int dk^D ~\delta(\omega_{12}-\omega)/(2\pi)^D$ is the JDOS. Eq.~\ref{sigma} provides an 
analytical formula for $\omega$ close to the band edge for a very general class of models. This simple
expression allows one to disentangle the contributions of the shift current integrand and 
the JDOS and hence to optimize them independently.

To maximize the response we therefore require band structures where the JDOS has a strong singularity. It is well known that in the 1D case, the generic JDOS diverges as a square root, $N(\omega) \propto (\omega - E_{\rm g})^{-1/2}$. 1D systems such as polymers or nanowires or systems in the quasi 1D limit will in general have a large response. In 2D, the band edge JDOS has a finite jump of $N(\omega) = (m_x m_y)^{1/2} /2\pi$, where $m_i$ are the average effective masses for valence and conduction bands. A singular $N(\omega)$ thus occurs in 2D when the inverse effective mass vanishes. In the effective model in Eq.~\ref{H2D}, this happens when $\delta=0$, which realizes what we may call a gapped semi-Dirac dispersion~\cite{BSP09}, since the coefficients of $\sigma_y$ and $\sigma_x$ are linear and quadratic in momentum, respectively. In such a case we have $N(\omega) \propto (\omega - E_{\rm g})^{-1/4}$  (full expressions for $N(\omega)$ may be found in the Methods section). 

For materials with large JDOS, the current can be further enhanced by appropriately tuning the parameters in Eqs.~\ref{eqn:integrandd_close_band_edge1}-\ref{eqn:integrandd_close_band_edge2}. This is most easily discussed if these parameters can be related to microscopic lattice models. In the next section, we discuss tight-binding models for simple materials that realize the described types of band structures. 

\begin{figure*}[t!]
\begin{center}
\includegraphics[width=18cm]{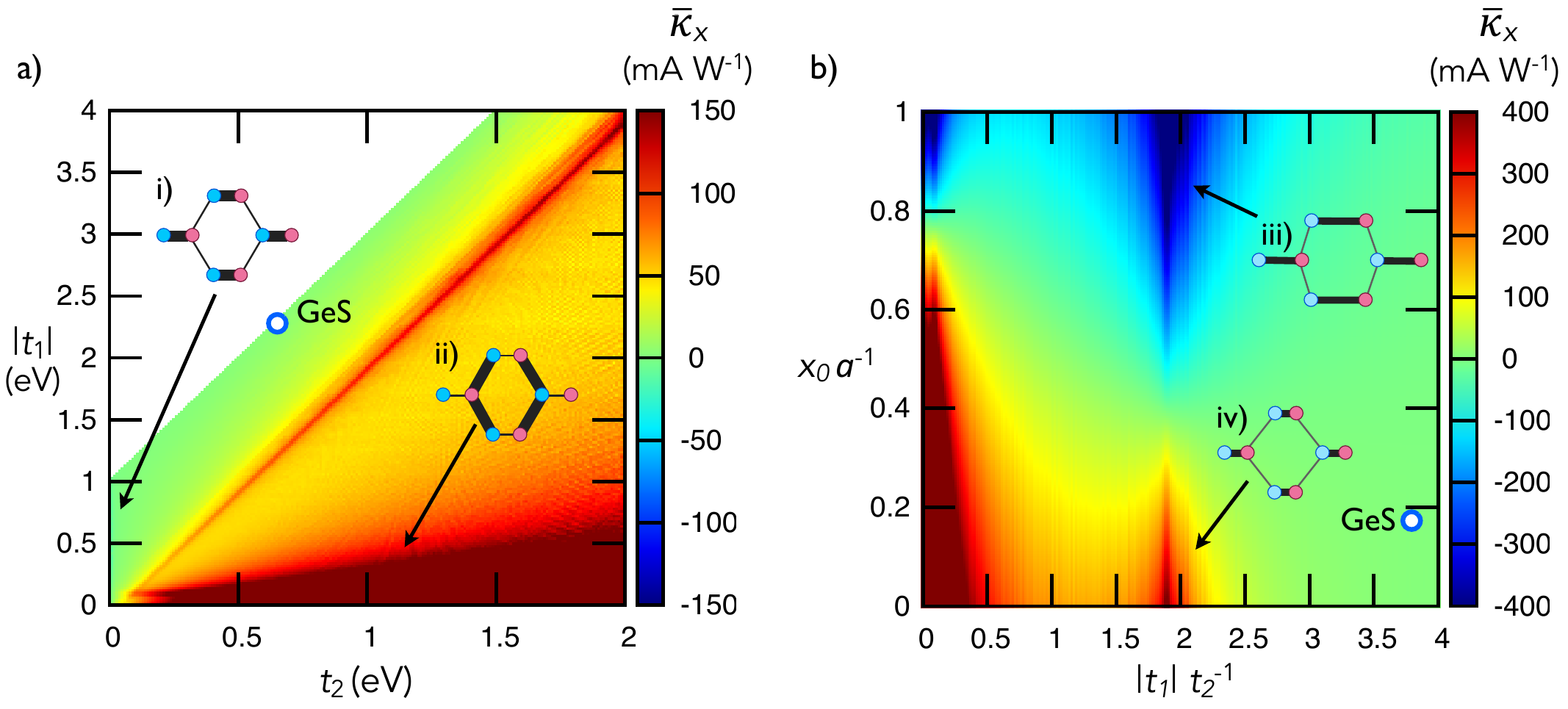}
\caption{Phase diagrams for monochalcogenide layer tight binding model: (a) Polarization-averaged photoresponsivity in the $x$-direction, $\bar{\kappa}_x$, at the band gap frequency plotted as a function of hopping parameters $|t_1|$ and $t_2$, keeping the band gap fixed at $1.89$ eV by tuning $\Delta$ accordingly. The Ge-S distance is $x_0=0.52$ ${\rm \AA}$ and $t_3=0$. The location of GeS on the phase diagram is marked by a white circle with blue outline. Regions for which the gap cannot be kept at $1.89$ eV are left white. (i) and (ii) show bond strengths in the limits where $|t_1| \gg t_2$ and $|t_1| \ll t_2$, respectively, to illustrate the two extremes of the phase diagram.  (b) Polarization-averaged photoresponsivity in the $x$-direction, $\bar{\kappa}_x$, at the band gap frequency plotted as a function of the Ge-S distance $x_0$ in units of $a=(\v{a}_1^2+\v{a}_2^2)^{1/2}$ and ratio of hopping parameters $|t_1| / t_2$.  Here, $\Delta$ and $t_2$ are set to GeS values of $1.1$ eV and $0.61$ eV, respectively.  The location of GeS on the phase diagram is marked by a white circle with blue outline.  (iii) and (iv) show two extreme cases of the phase diagram, where $x_0$ is large and small, respectively.} 
\label{phasediag}
\end{center}
\end{figure*}
\textbf{Material realizations and lattice models} - As a realization of the 1D case, we consider ferroelectric polymers that break inversion symmetry such as polyvinylidene fluoride or disubstituted polyacetilene\cite{SSH79,RM82,GFL99}. This system is described by the tight-binding model schematically shown in Fig.~\ref{fig1}(b), defined in terms of two types of hoppings, $t_1$ and $t_2$, alternating on-site potentials $\pm \Delta$, and orbital centers at $x=0$ and $x=x_0$. With our choice of basis functions, the Hamiltonian is specified by $f_x + i f_y = -[t_1 e^{i k_x x_0} + t_2 e^{-i k_x (a-x_0)}]$ and $f_z=\Delta$, where $a=10$ ${\rm \AA}$ is the lattice constant and the distance between closest neighbors is~\cite{SSH79} $x_0 = 0.48 a$. For estimates of the tight binding parameters, we consider the example of disubstituted polyacetilene that was experimentally realized in Ref.~\onlinecite{GFL99}, with a band gap of 2.5 eV.  For regular polyacetilene, where $\Delta=0$, the hopping parameters and band gap have been estimated as \cite{SSH79} $t_1 = 2.85$ eV, $t_2= 2.15$ eV, $E_{\rm g} = 1.4$ eV. Assuming the same hopping for the disubstituted version, we use $\Delta = 1.0$ eV to match the observed band gap. Note that the dispersion does not depend on $x_0$. 

Using Eq.~\ref{kappa_abb} and Eq.~\ref{intsum} we can now compute the shift current for this 1D model. Expanding about the low energy momentum $k_x = \pi/a$ and performing a constant rotation of the Pauli matrices, we obtain an effective model 
as Eq.~\ref{H2D} with parameters $k_y=0$ and  
$\delta = t_1 - t_2$, $v_F = (t_1-t_2)x_0 + t_2 a$, $\alpha_x = [t_2(a-x_0)^2-t_1x_0^2]/2$. 

To be able to compare the responsivity of these materials to that of a 3D system, we consider a stack of polymers as depicted in Fig.~\ref{fig1}(a), separated by a distance $d$ which we take to be equal to the lattice constant of the polymer $d=a$. The photoresponsivity is then $\kappa^{abb}_{3D} = \kappa^{abb}_{1D}/d^2$. The typical photoresponsivity spectrum of this model with this convention is shown in Fig.~\ref{fig:SC_vs_omega}(a).

For the 2D case, we require a layered material that breaks both inversion and rotational symmetries. The most popular of the recently isolated 2D semiconductors break either inversion (BN, MoS$_2$) or rotational symmetries (black phosphorus~\cite{XWJ14}, Re$S_2$ \cite{LFW15}), but not both. An inversion symmetry breaking version of the strongly anisotropic black phosphorus, a group V element, can be obtained combining elements of the IV and VI groups. These group IV monochalcogenides, such as GeS, are predicted to be stable in the monolayer form with the orthorhombic structure of black phosphorus~\cite{SH14,GC15}. 
 
These materials can be described with a tight binding model similar to the one used for black phosphorus\cite{RK14,RYK15,E14}. While the GeS unit cell contains two Ge-S pairs at different heights, a unit cell with a single Ge-S pair can be used when the physics to be probed is insensitive to the heights of the atoms (see Methods for a detailed explanation). The two band Hamiltonian is specified by $f_x + i f_y = -e^{-i \v{x}_0\cdot \v{k}}[t_1+t_2 \Phi(\v{k}) + t_3\Phi^*(\v{k})]$, where $\v{x}_0 = (x_0,0)$ and $\Phi(\v{k}) = (e^{i \v{a}_1\cdot \v{k}}+e^{i \v{a}_2\cdot \v{k}})$, and $f_z=\Delta$. $\v{a}_1$ and $\v{a}_2$ are the lattice vectors. See Fig.~\ref{fig1}(d) for the definition of the hopping integrals. Again note the dispersion is independent of $x_0$. The specific values of the tight-binding parameters for GeS have been obtained by fitting an ab-initio calculation as described in the Methods section, where the coefficients of the low energy model near the band edge are also shown. Note in this lattice structure there is a mirror symmetry $y\rightarrow -y$, which is represented as the identity, and restricts $\alpha_{xy} = \beta_{xy} = 0$. (This is so because both conduction and valence bands are even under the symmetry, as it also happens in black phosphorus. This is also the result of our ab-initio calculation.) This symmetry still allows a linear term of the form $k_x \sigma_y$, crucial for the semi-Dirac type of band structure. In this model, the semi-Dirac limit is realized when $t_1 = -2(t_2+t_3)$~\cite{MPF09}. 

We consider a stack of monolayers separated by $d=a$, as shown in Fig.~\ref{fig1}(c). In this case, we consider an inert spacer layer between the GeS layers to avoid the restoration of inversion symmetry that would occur if we were to stack GeS into its natural bulk form. The 3D photoresponsivity of this model, given by $\kappa^{abb}_{3D} = \kappa^{abb}_{2D}/d$, is computed using Eqs.~\ref{kappa_abb} and~\ref{intsum}. To make contact with the 1D case we consider a stacking distance $d = a \equiv (|\v{a}_1|+|\v{a}_2|)^{1/2}$ and $x_0 = 0.18 a$. The results are shown in Fig.~\ref{fig:SC_vs_omega}(b). We see that both $\kappa^{xxx}$ and $\kappa^{xyy}$ are in general finite, and the polarization average is also finite due to the strong anisotropy. 

The response of the monochalcogenides is large because they are close in parameter space to the gapped semi-Dirac Hamiltonian. This is best illustrated by considering the evolution of a fictitious system where the hoppings are tuned (with $t_3 =0$ for simplicity) to the semi-Dirac case $|t_1|/t_2=2$, where the divergence of the response is clearly appreciated. This evolution is shown in Fig.~\ref{fig:SC_vs_omega}(c).  

\textbf{Further optimization} - After describing the representative tight-binding models with large JDOS, we may now address a more systematic analysis of the photoresponsivity. First, we consider exploring the phase diagram of the monochalcogenides by sweeping $|t_1|, t_2$ in parameter space while the band gap is fixed at 1.89 eV by choosing $\Delta$ appropriately and $t_3=0$ for simplicity. Fig.~\ref{phasediag}(a) shows the polarization averaged photoresponsivity, $\bar{\kappa}_x = (\kappa^{xxx}+\kappa^{xyy})/2$, for the parameters $x_0=0.18a$ and $\theta = 0.69$. This phase diagram summarizes nicely the most physically relevant regimes where the shift current is large due to a divergent JDOS, namely the 1D dimensional limit where $|t_1| \ll t_2$, and the semi-Dirac regime where $|t_1| \sim 2 t_2$. In this phase diagram, the point corresponding to $t_1$ and $t_2$ of GeS is shown as a white circle with blue outline. 

Next we illustrate a very important feature of the behavior of the shift current integrand. Eqs.~\ref{eqn:integrandd_close_band_edge1}-\ref{eqn:integrandd_close_band_edge2} depend generically on the hoppings and lattice parameters. The energy does not depend on the parameter $x_0$, but the wavefunctions do. In Fig.~\ref{phasediag}(b), we show the peak photoresponsivity as a function of $|t_1|/t_2$ and $x_0$. A large response is observed in the semi-Dirac limit $|t_1|/t_2 \sim 2$. However, a very strong dependence on $x_0$ and even a sign change is also observed. The dependence on $x_0$ dramatically illustrates the fact that the shift current depends not only on the band structure, but also on the wavefunctions. This can be seen explicitly in the fact that the effective mass $m_x^{-1} =4a_x^2 t_1t_2/E_{\rm g}$ is independent of $x_0$, but the combination $v_F \alpha_x$ appearing in the shift current integrand is not. In particular $\alpha_x$ vanishes for $x_0 = a_x/[1+(|t_1/2t_2|)^{1/2}]$, which means that regardless of the JDOS, the band edge response can actually be zero. This behavior is characteristic of Berry connections, which depend explicitly on the positions of the sites in the unit cell. 

\textit{\textbf{Discussion}} - In this work, we have shown how an effective model for the band edge enables a clean separation of the two factors that contribute to a large shift current: the standard JDOS and the shift current matrix element. This model also allows us to readily identify materials with semi-Dirac-like Hamiltonians as those where both factors can be made large. Several other general conclusions can be drawn from the form of the effective shift current integrand in Eqs.~\ref{eqn:integrandd_close_band_edge1}-\ref{eqn:integrandd_close_band_edge2}. First, since the $1/\omega^3$ factor becomes $1/E_{\rm g}^3$ at the band edge, materials with smaller gaps are expected to have larger shift currents. A second conclusion is that while looking for materials with large JDOS is a good guiding principle, the shift current integrand depends on other microscopic details that can change the response dramatically. Within our simple model, the shift current can be maximized by bringing the two sites of the unit cell closer together, which is a requirement that the monochalcogenides satisfy well. Materials that may perform even better than GeS may be searched for exploring different chemical compositions, alloying, or by strain engineering. 

Our results were made possible by the derivation of a new sum rule appropriate for tight-binding models. With this sum rule, our work can be easily extended to tight-binding models with more than two bands, or systems where the minimum direct gap is not at a time-reversal invariant momentum. We expect that the formalism developed here will provide the necessary link to combine ab-initio methods with effective models, allowing for more in-depth, systematic study of shift current photovoltaics. 

Our results should be compared to known ferroelectric materials that have been recently studied. In the visible range of frequencies, $\omega \leq 3$ eV, we find peak values of 0.1 mAW$^{-1}$ in BiFeO$_3$ ~\cite{YZR12}, 1 mAW$^{-1}$ in hybrid perovskites~\cite{ZTW15} and a maximum 10 mAW$^{-1}$ in BaTiO$_3$\cite{YR12} or NaAsSe$_2$~\cite{BYZ14}. The realistic materials that we propose present larger responsivities, with the additional advantage that the peak is by construction at the band edge.  Moreover, as Fig. Fig.~\ref{fig:SC_vs_omega}(c) and Fig.~\ref{phasediag}(b) show, peak responses on the order of several hundreds of mAW$^{-1}$ could be achieved with materials closer to the semi-Dirac regime. To compare with conventional photovoltaic mechanisms, the total current per intensity of a crystalline Si solar cell exposed to sunlight is about 400 mAW$^{-1}$~\cite{flexible}. 

Given these numbers, our work is a sign that shift current photovoltaics capable of surpassing conventional solar cells may be close at hand, and a push to investigate their full potential using methods discussed in this work -- along with established techniques -- is warranted. We believe that the simple principles derived in our work will serve as a guide for both theory and experiment in the development and optimization of the next generation of shift current photovoltaics. 

\textit{\textbf{Methods}}

\textbf{Shift current} -  To make contact with previous work, we note the shift current integrand 
in Eq.~\ref{integrandd} is sometimes expressed in terms of the phase of the inter-band matrix 
element $r^b_{nm}= |r^b_{nm}| e^{i\phi_{nm}^b}$ as $I^{abb}_{nm} = |r^b_{nm}|^2 R^{a,b}_{nm}$ where

\begin{equation}
R^{a,b}_{nm} = \partial_{k_a}\phi_{nm}^b -\xi_{nn}^a+\xi_{mm}^a,
\end{equation}
is known as the shift vector. The response to a natural light source such as sunlight, which is unpolarized, is obtained by 
averaging $\kappa^{abb}$ over polarization. Taking $\vec E(\theta) = |E| (\cos \theta, \sin \theta)$ we have
\begin{equation}
\bar{J}_a = \int \frac{d\theta}{2\pi} J_a = \frac{1}{2}(\kappa^{axx}+\kappa^{ayy})I_0 = \bar{\kappa}_a I_0.
\label{unpolarized}
\end{equation}

\textbf{Sum rule} - 
The expression for the shift current presented in the main text can be obtained by the use of a 
sum rule for the quantity $r^a_{nm;b}$, which is obtained from the identity

\begin{equation}
\partial_{k_b}\partial_{k_a} \left<n|H|m\right> = \delta_{nm} \partial_{k_b}\partial_{k_a}E_n.
\end{equation}
Evaluating both sides explicitly for $n \neq m$, the identity can be expressed as 
\begin{align}
r^a_{nm;b}  = -\frac{1}{i\omega_{nm}} \left[ \frac{v^a_{nm}\Delta^b_{nm} +v^b_{nm}\Delta^a_{nm}}{\omega_{nm}} \right. \nonumber \\
 \left.-w_{nm}^{ab} + \sum_{p\neq n, m}(\frac{v^a_{np} v^b_{pm}}{\omega_{pm}}- \frac{v_{np}^b v^a_{pm}}{\omega_{np}})\right], && n \neq m
\end{align}
where $v^b_{nm} = \left<n|\partial_{k_b}H|m\right>$ are the velocity matrix elements, $\Delta^b_{nm} = v^b_{nn} - v^b_{mm}$, $w_{nm}^{ba}=\left<n|\partial_{k_b} \partial_{k_a}H|m\right>$ and $\omega_{nm} = E_n - E_m$. In the evaluation, we used 
\begin{align}
(r_{mn}^a)^* &= r^a_{nm}, \\
v^a_{nn} &=\partial_{k_a}E_n, \label{diag}\\
v^a_{nm} &=  i r^a_{nm}\omega_{nm}.  & n\neq m
\end{align}
The first equality follows from $\partial_k \left<n|m\right>= 0$ if $m \neq n$, while the last two follow from $\partial_{k_a} \left<n|H|m\right> = \delta_{nm} \partial_{k_a}E_n$. Note this sum rule contains the extra term $w_{nm}^{ab}$ compared to Ref.~\cite{Sipe2000}, where $H = p^2/2m +V(x)$ and $w_{nm}^{ab}=\delta_{nm}\delta^{ab}/m$ which has no off diagonal component. Quite importantly, the term $w_{nm}^{ab}$ in tight binding models is the one responsible for all band edge contributions. Also note that it has been argued before that $I^{xxx}=0$ for a two band model~\cite{Baltz1981}, which is actually only true if $w_{nm}^{ab}=0$.

\textbf{Two band model} - For the case of two bands, $m=1$, $n=2$ the use of the sum rule for the shift current integrand in Eq.~\ref{integrandd} leads to the simplified expression 
\begin{equation}
I^{abb}_{nm} = \frac{1}{\omega_{12}^2}{\rm Im} \left[\frac{-v^b_{21}v^a_{12}(v^b_{11}-v^b_{22})}{2 \epsilon} +v^b_{21}w_{12}^{ba}\right]. \label{twobandint}
\end{equation}
To evaluate this expression we compute the wave functions of $H$
\begin{align}
\psi_{n} =  \frac{1}{\sqrt{2\epsilon}}(-\eta \sqrt{\epsilon -\eta f_z} \:, e^{i \phi_{\v{k}} }\sqrt{\epsilon +\eta f_z}),
\end{align}
with $n=1,2$, $\eta=(-1)^n$, and $\phi_{\v{k}} = \arctan (f_y/f_x)$. The required matrix elements are
\begin{align}
v_{21}^a &= \big< \psi_2| (\epsilon_{0,a}  \mathcal{I} +  \sum_i\sigma_i f_{i,a} |\psi_1\big> =  \sum_i f_{i,a}s_i^*, \\
w_{12}^{ab} &= \big< \psi_1| (\epsilon_{0,ab}  \mathcal{I} +  \sum_i \sigma_i f_{i,ab} |\psi_2\big> =  \sum_i f_{i,ab}s_i,
\end{align}
where the off diagonal matrix element $s_{i} = \left< \psi_1| \sigma_i |\psi_2\right>$ is
\begin{align}
s_i = \big(\tfrac{f_z}{\epsilon} \cos \phi_{\v{k}} + i \sin \phi_{\v{k}}, \tfrac{f_z}{\epsilon} \sin \phi_{\v{k}} \nonumber 
\\ - i\cos \phi_{\v{k}},-\tfrac{(f_x^2+f_y^2)^{1/2}}{\epsilon}\big), 
\end{align}
and the diagonal velocity matrix elements are computed from Eq.~\ref{diag}. The imaginary part in 
Eq.~\ref{twobandint} can be taken using ${\rm Im} \left[ s_i^* s_j\right] = -\sum_{m}{\epsilon_{ijm} f_m/\epsilon}$
and this leads to Eq.~\ref{intsum} in the main text. 

\textbf{Joint density of states} - To compute the JDOS, we first start with the 1D case. Close to the band edge, we expand the energies of conduction and valence bands as $E_i \approx E_i(0) + k_x^2/2m_{i,x} $, so that $\omega_{12} = E_1-E_2 \approx E_{\rm g} + k_x^2/2 m_x$ where the total effective mass $m_x^{-1} = |m_{1,x}|^{-1}+|m_{2,x}|^{-1}$ is given by
\begin{equation}
m_x^{-1}=4(v_F^2+2\alpha_x \delta +2 \beta_x \Delta)/E_{\rm g} ,\label{zx}
\end{equation}
and solve for $k(\omega) = \sqrt{2m_x(\omega-E_{\rm g})}$. Rescaling $2m_x$ we get

\begin{align}
N^{1D}(\omega) &= \sqrt{2m_x} \int \frac{dk}{2\pi} \frac{\delta(k\pm k(\omega))}{|2 k|} \nonumber \\
 &= \frac{\sqrt{2m_x} }{2\pi}\frac{\theta(\omega-E_{\rm g})}{\sqrt{(\omega - E_{\rm g})}},
\end{align}
where we get the expected 1D singularity. For the generic 2D case, again we expand 
$\omega_{12} \approx E_{\rm g} + k_x^2/2m_x +k_y^2/2m_y$, where $m_x$ is still given by Eq.~\ref{zx} and
\begin{align}
m_y^{-1} = 8(\alpha_y \delta +\beta_y \Delta)/E_{\rm g},
\end{align}
We consider the case when $m_x>0$, $m_y>0$, so that the minimum does lie at $\vec k=0$. 
By rescaling $2m_{x}$ and $2m_y$ we get in polar coordinates
\begin{align}
N^{2D} &= \sqrt{4 m_x m_y}\int \frac{kdk d\theta}{(2\pi)^2} \frac{\delta(k-k(\omega))}{|2k|}  \nonumber \\ &=  \frac{\sqrt{m_x m_y}}{2\pi} \theta(\omega-E_{\rm g}),
\end{align}
which is the expected constant result. Finally, the semi-Dirac case occurs in 2D when $m_y^{-1}=0$, 
which in the absence of second neighbor hopping occurs exactly at $\delta=0$. 
In this case, we keep the complete expression for $\omega_{12} = ((\alpha_x k_x^2+ \alpha_y k_y^2)^2+v_F^2k_x^2+\Delta^2)^{1/2}$. 
In polar coordinates we have

\begin{equation}
N^{SD} =\int \frac{kdk d\theta}{(2\pi)^2} \frac{\delta(k-k(\omega))}{|\partial_k \omega_{12}|}.
\end{equation}
We now rescale $\alpha_{x}$, $\alpha_y$ instead, solve for $k$

\begin{equation}
k(\omega) = [-v_F^2/\alpha_x \cos^2 \theta \pm (v_F^4/\alpha_x^2 \cos ^4\theta +\omega^2-E_{\rm g}^2)^{1/2}]/2 ,\nonumber
\end{equation}
and get
\begin{align}
 N^{SD} &= \frac{\omega}{4\sqrt{\alpha_x\alpha_y}}\int \frac{dk d\theta}{(2\pi)^2} \frac{\delta(k-k(\omega))}{(v_F^4/\alpha_x^2 \cos^4 \theta+\omega^2-E_{\rm g}^2)^{1/2}} \nonumber \\
 &= \frac{\Gamma(\tfrac{1}{4})}{4\Gamma(\tfrac{3}{4})(2\pi)^{3/2}|\alpha_x|\sqrt{\alpha_y}v_F} \frac{\omega \theta(\omega-E_{\rm g})}{(\omega^2-E_{\rm g}^2)^{1/4}}.
\end{align}

\begin{figure}[h]
\begin{center}
\includegraphics[width=6cm]{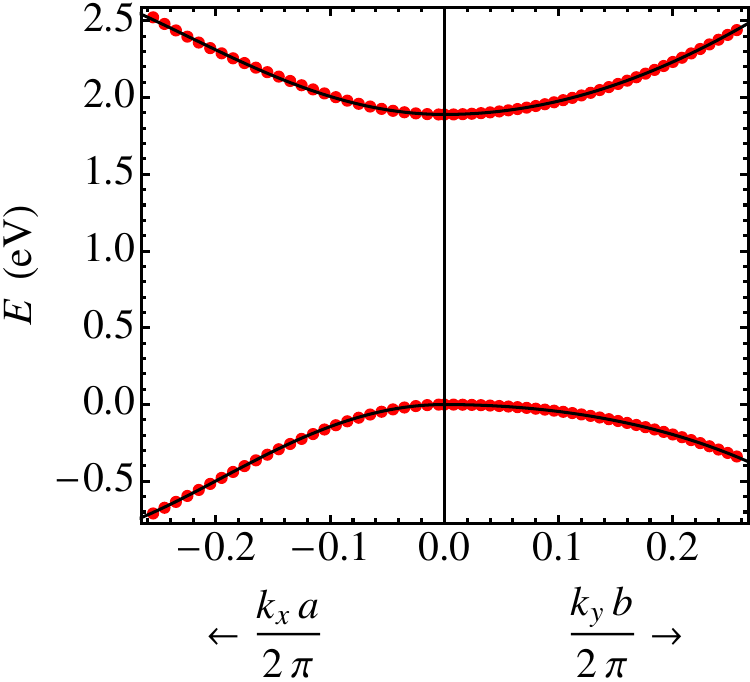}
\caption{Tight binding fit to ab initio for GeS: Dispersion of conduction and valence bands of GeS near $\Gamma$ computed ab-initio (red dots). A black line shows the tight binding fit for comparison.} 
\label{schematic}
\end{center}
\end{figure}

\textbf{Ab-initio calculation and tight binding fit for GeS} -
Due to the lack of tight binding models for monochalcogenide materials~\cite{SH14,GC15}, we have 
derived the tight binding parameters by fitting the electronic structure of GeS ab-initio. 
We used the PBE~\cite{pbe1997} approximation to the exchange correlation functional, ultrasoft 
pseudopotentials,~\cite{Garrity2014} Quantum-ESPRESSO~\cite{giannozzi} and Wannier90~\cite{Mostofi2008} 
computer packages.  The cutoff for electron wavefunction is set to 40 Ry and cutoff for electron 
density to 200 Ry. Internal coordinates and in-plane lattice constants were fully relaxed.  
Vacuum region between repeating images of GeS monolayers is 17 ${\rm \AA}$.  Wannier functions 
were constructed from a 12x12 regular k-mesh grid. The maximally localized Wannier functions were constructed in a standard way by projecting into hydrogenic s-like and p-like orbitals on both Ge and S atoms along with two s-like orbitals in the vacuum region that are needed to represent the vacuum states.  The frozen window for the disentanglement procedure spans up to 6.2 eV above the Fermi level.
The crystal structure of GeS is orthorombic with space group Pnma (No. 62) and
lattice vectors $\vec l_1 = (l_1,0)$ and $\vec l_2 = (0,l_2)$, with $l_1 = 4.53 \;{\rm \AA}$ 
and $l_2 = 3.63 \;{\rm \AA}$ and contains two Ge and two S atoms. The structure can be seen as two 
GeS zigzag chains separated by a height of $h = 2.32\;{\rm \AA}$. The ab-initio results for the 
conduction and valence bands near the $\Gamma$ point are shown in Fig. \ref{schematic} and have 
mostly $p_z$ character. 

This system can be effectively described with a two site tight binding model. This can be done 
because the lattice structure has glide symmetries with mirror reflection $z \rightarrow -z$ and 
translations $\vec a_1 = (a_x,a_y)$ and $\vec a_2 = (a_x,-a_y)$, with $a_x = l_1/2$ and $a_y = l_2/2$. 
When the out of plane positions of the atoms are not relevant for the problem of interest, one can 
define a smaller two site unit cell where the glides play the role of lattice vectors (as it is 
done in black phosphorus \cite{E14}). The Ge and S sites in this effective tight binding model 
are located at $(0,0)$ and $(x_0,0)$, with $x_0 = 0.62 \;{\rm \AA}$.This is the tight binding model 
employed in the main text. The parameters of this model are obtained from the ab-initio calculation as follows. 

Since our aim is to model faithfully only the low energy bands around the Gamma point, it will 
suffice to consider a single $p_z$ orbital per site in the tight binding model. The minimal 
model parameters are the on-site potential difference $\Delta$ between Ge and S $p_z$ orbitals 
and the three nearest neighbors hoppings $t_i$, with $i=1,2,3$, which are all between Ge and S 
atoms. In addition, to reproduce the small particle-hole asymmetry of the gap, we also consider 
two further neighbor hoppings $t_1'$ and $t_2'$ which connect Ge-Ge or S-S pairs (we assume the 
same values for both species to simplify). 

The tight binding Hamiltonian takes the form $H = \epsilon_0 + \Sigma_i \sigma_i f_i(\v{k})$ with coefficients
\begin{align}
\epsilon_0 =& -2t_1'(\cos \v{a}_1\cdot \v{k} +\cos \v{a}_2\cdot \v{k}) \nonumber\\
 &- 2t_2' \cos (\v{a}_1-\v{a}_2)\cdot \v{k},\\
f_x + if_y =& -e^{-i\v{x}_0\cdot \v{k}}[t_1 + t_2 \Phi(\v{k})+t_3 \Phi^*(\v{k})],\\
f_z =& \Delta ,
\end{align}
where, as defined in the text, $\Phi(\v{k}) = (e^{i \v{a}_1\cdot \v{k}}+e^{i \v{a}_2 \cdot \v{k}})$. Our tight binding fit is intended to reproduce faithfully the bands and wavefunctions close to the band edge, where the effective low energy model applies. This model is given by
\begin{align}
H &= (\gamma_x k_x^2+\gamma_y k_y^2)\mathcal{I} +(\delta + \alpha_x k_x^2 + \alpha_y k_y^2)\sigma_x  \nonumber \\
&+ v_F k_x \sigma_y + \Delta \sigma_z ,
\end{align}
where a constant term is omitted as it can be absorbed in the chemical potential. The effective model parameters are related to the tight binding parameters as
\begin{align}
\gamma_x &= 2t_1'a_x^2,\\
\gamma_y &= (2t_1'+4t_2')a_y^2,\\
\delta &= t_1 - 2t_2 -2t_3, \\
v_F &= -2a_x (t_2 - t_3) - (t_1 - 2 t_2 -2 t_3) x_0, \\
\alpha_x &= t_2 (a_x - x_0)^2 - t_1 x_0^2/2 + t_3 (a_x + x_0)^2,\\
\alpha_y &=  (t_2+t_3) a_y^2.
\end{align}
\
The key to obtain a reliable tight binding parametrization is that, since the shift current depends sensitively on the actual wavefunctions, the tight binding model should be fitted to wavefunction dependent quantities in addition to the band energies. The simplest gauge invariant quantity that depends on wavefunction phases is the bracket of two covariant derivatives 
\begin{equation}
Q_{\mu\nu} = \left< D_\mu u_k | D_\nu u_k\right>,
\end{equation}
with $D_\mu = \partial_{\mu} - i A_\mu$, with $A_\mu = i \left<u_k|\partial_\mu u_k \right>$ the Berry connection. The real and imaginary parts of this tensor are known as the Berry curvature and the quantum metric. A fit that reproduces this tensor correctly in addition to band energies ensures that the wavefunction structure around the $\Gamma$ point is correctly accounted for, so that any other gauge invariant quantity computed in the effective model should be the same as that computed ab-initio. 

\begin{center}
\begin{table*}[t]
\begin{tabular}{|c|c|c|c|c|c|c|}
\hline
  \multicolumn{7}{|c|}{Ab-initio input parameters}  \\ \hline
$E_{\rm g}$ & $m_{x,v}$ & $m_{x,c}$ & $m_{y,v}$ & $m_{y,c}$ & $\partial_y \Omega$ & $g_{xx}$ \\ \hline
  1.89 eV & -0.064 ${\rm{eV^{-1} \AA^{-2}}}$ & 0.079 ${\rm{eV^{-1} \AA^{-2}}}$ & -0.340 ${\rm{eV^{-1} \AA^{-2}}}$ & 0.171 ${\rm{eV^{-1} \AA^{-2}}}$ & 3.565 ${\rm{\AA^{3}}}$ & 2.529 $\rm{\AA^{2}}$ \\ \hline
  \multicolumn{7}{|c|}{Tight binding parameters}  \\ \hline
$\Delta$ & $t_1$ & $t_2$  & $t_3$ & $t_1'$ & $t_2'$  & $x_0$ \\ \hline
0.41 eV & -2.33 eV & 0.61 eV & 0.13 eV & 0.07 eV  & -0.09 eV & 0.52 $\rm{\AA}$ \\ \hline
\end{tabular}
\caption{Table of ab-initio and tight binding parameters for GeS: First row: input ab-initio parameters. Second row: Tight binding parameters obtained from 
the fitting. }\label{tab}
\end{table*}
\end{center}

The Berry curvature $\Omega(k)$ is defined as
\begin{equation}
\Omega(k) = \epsilon_{\mu\nu} {\rm Im} [ \left<\partial_\mu u_k| \partial_\nu u_k \right>] = \nabla \times A.
\end{equation}
The Berry curvature around $\Gamma$ for the tight binding model is given by
\begin{equation}
\Omega = \frac{v_F (\alpha_y \Delta-\beta_y \delta)}{(\Delta^2+\delta^2)^{3/2}}k_y.
\end{equation}
Since $\Omega$ vanishes at the origin, we take $\partial_y \Omega$ as one extra input for the fit. The quantum metric is defined as
\begin{equation}
g_{\mu\nu} = {\rm Re} [ \left<\partial_\mu u_k| \partial_\nu u_k \right>] - A_\mu A_\nu. \label{metric}
\end{equation}
The only non-vanishing component of the quantum metric at $k=0$ is given by
\begin{align}
g_{xx} &= \frac{v_F^2}{4(\Delta^2+\delta^2)},
\end{align}
so we take $g_{xx}$ as another extra input for the fit. 

In summary, we take as ab-initio input parameters the gap, the four effective masses, and the lowest order Berry curvature and quantum metric, $\partial_y \Omega$ and $g_{xx}$. The difference in effective masses for electron and hole bands, accounted for the term $\epsilon_0$, can be fitted independently with the hoppings $t_1'$ and $t_2'$. Since $\epsilon_0$ has no impact in the shift current response, the hoppings $t_1'$ and $t_2'$ are not considered in the main text. The rest of the input is fitted with $t_1$, $t_2$ and $t_3$, the on-site potential $\Delta$ and $x_0$, and the results of the fit are shown in Table \ref{tab}. While $x_0$ is in fact known from the lattice structure of GeS to be $0.62 \rm{\AA}$, obtaining it independently from the tight binding fit, which gives a close value of $0.52 \rm{\AA}$ provides an additional check of the validity of the model. 

\textbf{Acknowledgments} - We acknowledge useful discussions with J. Sipe, E. J. Mele, M. Bernardi, P. Kr\'al, S. Barraza-Lopez and F. Duque-Gomez and especially with Y. Xu. We also thank R. Ilan and 
A.G. Grushin for a careful reading of the manuscript. BMF was supported by Conacyt, NSF 
DMR-12065135 and NERSC Contract No. DE-AC02-05CH11231, AMC was supported by the NSERC CGS-MSFSS and the NSERC CGS-D3, F. de Juan was supported by the U.S. Department of Energy, Office of Science, Basic Energy Sciences, Materials Sciences and Engineering Division, grant DE-AC02-05CH11231, and J.E.M. was supported by AFOSR MURI.

\textbf{Author contributions} - AMC and BMF contributed equally to the work. AMC, BMF, FJ and JEM carried out 
the analytical and numerical analysis. SC carried out all ab-initio computations. All authors contributed 
to the results and the writing of the manuscript.

\textbf{Data availability} - The data that support the findings of this study are available from the corresponding author upon request.

\textbf{Competing financial interests} - The authors declare no competing financial interests.

\textbf{Correspondence} - Correspondence and requests for materials should be addressed to J.E.M. (email: jemoore@berkeley.edu)

\bibliography{shift_current}

\begin{thebibliography}{74}%
\makeatletter
\providecommand \@ifxundefined [1]{%
 \@ifx{#1\undefined}
}%
\providecommand \@ifnum [1]{%
 \ifnum #1\expandafter \@firstoftwo
 \else \expandafter \@secondoftwo
 \fi
}%
\providecommand \@ifx [1]{%
 \ifx #1\expandafter \@firstoftwo
 \else \expandafter \@secondoftwo
 \fi
}%
\providecommand \natexlab [1]{#1}%
\providecommand \enquote  [1]{``#1''}%
\providecommand \bibnamefont  [1]{#1}%
\providecommand \bibfnamefont [1]{#1}%
\providecommand \citenamefont [1]{#1}%
\providecommand \href@noop [0]{\@secondoftwo}%
\providecommand \href [0]{\begingroup \@sanitize@url \@href}%
\providecommand \@href[1]{\@@startlink{#1}\@@href}%
\providecommand \@@href[1]{\endgroup#1\@@endlink}%
\providecommand \@sanitize@url [0]{\catcode `\\12\catcode `\$12\catcode
  `\&12\catcode `\#12\catcode `\^12\catcode `\_12\catcode `\%12\relax}%
\providecommand \@@startlink[1]{}%
\providecommand \@@endlink[0]{}%
\providecommand \url  [0]{\begingroup\@sanitize@url \@url }%
\providecommand \@url [1]{\endgroup\@href {#1}{\urlprefix }}%
\providecommand \urlprefix  [0]{URL }%
\providecommand \Eprint [0]{\href }%
\providecommand \doibase [0]{http://dx.doi.org/}%
\providecommand \selectlanguage [0]{\@gobble}%
\providecommand \bibinfo  [0]{\@secondoftwo}%
\providecommand \bibfield  [0]{\@secondoftwo}%
\providecommand \translation [1]{[#1]}%
\providecommand \BibitemOpen [0]{}%
\providecommand \bibitemStop [0]{}%
\providecommand \bibitemNoStop [0]{.\EOS\space}%
\providecommand \EOS [0]{\spacefactor3000\relax}%
\providecommand \BibitemShut  [1]{\csname bibitem#1\endcsname}%
\let\auto@bib@innerbib\@empty
\bibitem [{\citenamefont {Shockley}\ and\ \citenamefont
  {Queisser}(1961)}]{Shockley1961}%
  \BibitemOpen
  \bibfield  {author} {\bibinfo {author} {\bibfnamefont {William}\ \bibnamefont
  {Shockley}}\ and\ \bibinfo {author} {\bibfnamefont {Hans~J.}\ \bibnamefont
  {Queisser}},\ }\bibfield  {title} {\enquote {\bibinfo {title} {Detailed
  balance limit of efficiency of pn junction solar cells},}\ }\href@noop {}
  {\bibfield  {journal} {\bibinfo  {journal} {J. Appl. Phys.}\ }\textbf
  {\bibinfo {volume} {32}},\ \bibinfo {pages} {510} (\bibinfo {year}
  {1961})}\BibitemShut {NoStop}%
\bibitem [{\citenamefont {Kraut}\ and\ \citenamefont {von Baltz}(1979)}]{KB79}%
  \BibitemOpen
  \bibfield  {author} {\bibinfo {author} {\bibfnamefont {Wolfgang}\
  \bibnamefont {Kraut}}\ and\ \bibinfo {author} {\bibfnamefont {Ralph}\
  \bibnamefont {von Baltz}},\ }\bibfield  {title} {\enquote {\bibinfo {title}
  {Anomalous bulk photovoltaic effect in ferroelectrics: A quadratic response
  theory},}\ }\href {\doibase 10.1103/PhysRevB.19.1548} {\bibfield  {journal}
  {\bibinfo  {journal} {Phys. Rev. B}\ }\textbf {\bibinfo {volume} {19}},\
  \bibinfo {pages} {1548--1554} (\bibinfo {year} {1979})}\BibitemShut {NoStop}%
\bibitem [{\citenamefont {Belinicher}\ and\ \citenamefont
  {Sturman}(1980)}]{BS80}%
  \BibitemOpen
  \bibfield  {author} {\bibinfo {author} {\bibfnamefont {V~I}\ \bibnamefont
  {Belinicher}}\ and\ \bibinfo {author} {\bibfnamefont {B~I}\ \bibnamefont
  {Sturman}},\ }\bibfield  {title} {\enquote {\bibinfo {title} {The
  photogalvanic effect in media lacking a center of symmetry},}\ }\href
  {http://stacks.iop.org/0038-5670/23/i=3/a=R02} {\bibfield  {journal}
  {\bibinfo  {journal} {Sov. Phys. Usp.}\ }\textbf {\bibinfo {volume} {23}},\
  \bibinfo {pages} {199} (\bibinfo {year} {1980})}\BibitemShut {NoStop}%
\bibitem [{\citenamefont {von Baltz}\ and\ \citenamefont
  {Kraut}(1981)}]{Baltz1981}%
  \BibitemOpen
  \bibfield  {author} {\bibinfo {author} {\bibfnamefont {Ralph}\ \bibnamefont
  {von Baltz}}\ and\ \bibinfo {author} {\bibfnamefont {Wolfgang}\ \bibnamefont
  {Kraut}},\ }\bibfield  {title} {\enquote {\bibinfo {title} {Theory of the
  bulk photovoltaic effect in pure crystals},}\ }\href@noop {} {\bibfield
  {journal} {\bibinfo  {journal} {Phys. Rev. B}\ }\textbf {\bibinfo {volume}
  {23}},\ \bibinfo {pages} {5590} (\bibinfo {year} {1981})}\BibitemShut
  {NoStop}%
\bibitem [{\citenamefont {Presting}\ and\ \citenamefont
  {Von~Baltz}(1982)}]{PB82}%
  \BibitemOpen
  \bibfield  {author} {\bibinfo {author} {\bibfnamefont {H.}~\bibnamefont
  {Presting}}\ and\ \bibinfo {author} {\bibfnamefont {R.}~\bibnamefont
  {Von~Baltz}},\ }\bibfield  {title} {\enquote {\bibinfo {title} {Bulk
  photovoltaic effect in a ferroelectric crystal a model calculation},}\ }\href
  {\doibase 10.1002/pssb.2221120225} {\bibfield  {journal} {\bibinfo  {journal}
  {Phys. Status Solidi (b)}\ }\textbf {\bibinfo {volume} {112}},\ \bibinfo
  {pages} {559--564} (\bibinfo {year} {1982})}\BibitemShut {NoStop}%
\bibitem [{\citenamefont {Sturman}\ and\ \citenamefont
  {Sturman}(1992)}]{Sturman1992}%
  \BibitemOpen
  \bibfield  {author} {\bibinfo {author} {\bibfnamefont {Boris~I.}\
  \bibnamefont {Sturman}}\ and\ \bibinfo {author} {\bibfnamefont {Paul~J.}\
  \bibnamefont {Sturman}},\ }\href@noop {} {\emph {\bibinfo {title}
  {Photovoltaic and Photo-refractive Effects in Noncentrosymmetric
  Materials}}}\ (\bibinfo  {publisher} {CRC Press},\ \bibinfo {year}
  {1992})\BibitemShut {NoStop}%
\bibitem [{\citenamefont {Aversa}\ and\ \citenamefont {Sipe}(1995)}]{AS95}%
  \BibitemOpen
  \bibfield  {author} {\bibinfo {author} {\bibfnamefont {Claudio}\ \bibnamefont
  {Aversa}}\ and\ \bibinfo {author} {\bibfnamefont {J.~E.}\ \bibnamefont
  {Sipe}},\ }\bibfield  {title} {\enquote {\bibinfo {title} {Nonlinear optical
  susceptibilities of semiconductors: Results with a length-gauge analysis},}\
  }\href {\doibase 10.1103/PhysRevB.52.14636} {\bibfield  {journal} {\bibinfo
  {journal} {Phys. Rev. B}\ }\textbf {\bibinfo {volume} {52}},\ \bibinfo
  {pages} {14636--14645} (\bibinfo {year} {1995})}\BibitemShut {NoStop}%
\bibitem [{\citenamefont {Kristoffel}\ \emph {et~al.}(1982)\citenamefont
  {Kristoffel}, \citenamefont {von Baltz},\ and\ \citenamefont
  {Hornung}}]{KBH82}%
  \BibitemOpen
  \bibfield  {author} {\bibinfo {author} {\bibfnamefont {N.}~\bibnamefont
  {Kristoffel}}, \bibinfo {author} {\bibfnamefont {R.}~\bibnamefont {von
  Baltz}}, \ and\ \bibinfo {author} {\bibfnamefont {D.}~\bibnamefont
  {Hornung}},\ }\bibfield  {title} {\enquote {\bibinfo {title} {On the
  intrinsic bulk photovoltaic effect: Performing the sum over intermediate
  states},}\ }\href {\doibase 10.1007/BF01313794} {\bibfield  {journal}
  {\bibinfo  {journal} {Z. Physik}\ }\textbf {\bibinfo {volume} {47}},\
  \bibinfo {pages} {293--296} (\bibinfo {year} {1982})}\BibitemShut {NoStop}%
\bibitem [{\citenamefont {Sipe}\ and\ \citenamefont
  {Shkrebtii}(2000)}]{Sipe2000}%
  \BibitemOpen
  \bibfield  {author} {\bibinfo {author} {\bibfnamefont {J.~E.}\ \bibnamefont
  {Sipe}}\ and\ \bibinfo {author} {\bibfnamefont {A.~I.}\ \bibnamefont
  {Shkrebtii}},\ }\bibfield  {title} {\enquote {\bibinfo {title} {Second-order
  optical response in semiconductors},}\ }\href@noop {} {\bibfield  {journal}
  {\bibinfo  {journal} {Phys. Rev. B}\ }\textbf {\bibinfo {volume} {61}},\
  \bibinfo {pages} {5337} (\bibinfo {year} {2000})}\BibitemShut {NoStop}%
\bibitem [{\citenamefont {Kr\'al}\ \emph {et~al.}(2000)\citenamefont {Kr\'al},
  \citenamefont {Mele},\ and\ \citenamefont {Tom\'anek}}]{KMT00}%
  \BibitemOpen
  \bibfield  {author} {\bibinfo {author} {\bibfnamefont {Petr}\ \bibnamefont
  {Kr\'al}}, \bibinfo {author} {\bibfnamefont {E.~J.}\ \bibnamefont {Mele}}, \
  and\ \bibinfo {author} {\bibfnamefont {David}\ \bibnamefont {Tom\'anek}},\
  }\bibfield  {title} {\enquote {\bibinfo {title} {Photogalvanic effects in
  heteropolar nanotubes},}\ }\href@noop {} {\bibfield  {journal} {\bibinfo
  {journal} {Phys. Rev. Lett.}\ }\textbf {\bibinfo {volume} {85}},\ \bibinfo
  {pages} {1512} (\bibinfo {year} {2000})}\BibitemShut {NoStop}%
\bibitem [{\citenamefont {Ji}\ \emph {et~al.}(2010)\citenamefont {Ji},
  \citenamefont {Yao},\ and\ \citenamefont {Liang}}]{bpve_abovebgvoltage}%
  \BibitemOpen
  \bibfield  {author} {\bibinfo {author} {\bibfnamefont {Wei}\ \bibnamefont
  {Ji}}, \bibinfo {author} {\bibfnamefont {Kui}\ \bibnamefont {Yao}}, \ and\
  \bibinfo {author} {\bibfnamefont {Yung~C.}\ \bibnamefont {Liang}},\
  }\bibfield  {title} {\enquote {\bibinfo {title} {Bulk photovoltaic effect at
  visible wavelength in epitaxial ferroelectric bifeo3 thin films},}\ }\href
  {\doibase 10.1002/adma.200902985} {\bibfield  {journal} {\bibinfo  {journal}
  {Adv. Mater.}\ }\textbf {\bibinfo {volume} {22}},\ \bibinfo {pages}
  {1763--1766} (\bibinfo {year} {2010})}\BibitemShut {NoStop}%
\bibitem [{\citenamefont {Zheng}\ \emph {et~al.}(2015)\citenamefont {Zheng},
  \citenamefont {Takenaka}, \citenamefont {Wang}, \citenamefont {Koocher},\
  and\ \citenamefont {Rappe}}]{ZTW15}%
  \BibitemOpen
  \bibfield  {author} {\bibinfo {author} {\bibfnamefont {Fan}\ \bibnamefont
  {Zheng}}, \bibinfo {author} {\bibfnamefont {Hiroyuki}\ \bibnamefont
  {Takenaka}}, \bibinfo {author} {\bibfnamefont {Fenggong}\ \bibnamefont
  {Wang}}, \bibinfo {author} {\bibfnamefont {Nathan~Z.}\ \bibnamefont
  {Koocher}}, \ and\ \bibinfo {author} {\bibfnamefont {Andrew~M.}\ \bibnamefont
  {Rappe}},\ }\bibfield  {title} {\enquote {\bibinfo {title} {First-principles
  calculation of the bulk photovoltaic effect in ch3nh3pbi3 and
  ch3nh3pbi3â€“xclx},}\ }\href@noop {} {\bibfield  {journal} {\bibinfo
  {journal} {J. Phys. Chem. Lett.}\ }\textbf {\bibinfo {volume} {6}},\ \bibinfo
  {pages} {31--37} (\bibinfo {year} {2015})}\BibitemShut {NoStop}%
\bibitem [{\citenamefont {Young}\ and\ \citenamefont {Rappe}(2012)}]{YR12}%
  \BibitemOpen
  \bibfield  {author} {\bibinfo {author} {\bibfnamefont {Steve~M.}\
  \bibnamefont {Young}}\ and\ \bibinfo {author} {\bibfnamefont {Andrew~M.}\
  \bibnamefont {Rappe}},\ }\bibfield  {title} {\enquote {\bibinfo {title}
  {First principles calculation of the shift current photovoltaic effect in
  ferroelectrics},}\ }\href@noop {} {\bibfield  {journal} {\bibinfo  {journal}
  {Phys. Rev. Lett.}\ }\textbf {\bibinfo {volume} {109}},\ \bibinfo {pages}
  {116601} (\bibinfo {year} {2012})}\BibitemShut {NoStop}%
\bibitem [{\citenamefont {Brehm}\ \emph {et~al.}(2014)\citenamefont {Brehm},
  \citenamefont {Young}, \citenamefont {Zheng},\ and\ \citenamefont
  {Rappe}}]{BYZ14}%
  \BibitemOpen
  \bibfield  {author} {\bibinfo {author} {\bibfnamefont {John~A}\ \bibnamefont
  {Brehm}}, \bibinfo {author} {\bibfnamefont {Steve~M}\ \bibnamefont {Young}},
  \bibinfo {author} {\bibfnamefont {Fan}\ \bibnamefont {Zheng}}, \ and\
  \bibinfo {author} {\bibfnamefont {Andrew~M}\ \bibnamefont {Rappe}},\
  }\bibfield  {title} {\enquote {\bibinfo {title} {First-principles calculation
  of the bulk photovoltaic effect in the polar compounds liass2, liasse2, and
  naasse2},}\ }\href@noop {} {\bibfield  {journal} {\bibinfo  {journal} {J.
  Chem. Phys.}\ }\textbf {\bibinfo {volume} {141}},\ \bibinfo {pages} {204704}
  (\bibinfo {year} {2014})}\BibitemShut {NoStop}%
\bibitem [{\citenamefont {Hodes}(2013)}]{Hodes2013}%
  \BibitemOpen
  \bibfield  {author} {\bibinfo {author} {\bibfnamefont {Gary}\ \bibnamefont
  {Hodes}},\ }\bibfield  {title} {\enquote {\bibinfo {title} {Perovskite-based
  solar cells},}\ }\href@noop {} {\bibfield  {journal} {\bibinfo  {journal}
  {Science}\ }\textbf {\bibinfo {volume} {342}},\ \bibinfo {pages} {317}
  (\bibinfo {year} {2013})}\BibitemShut {NoStop}%
\bibitem [{\citenamefont {Egger}\ \emph {et~al.}(2015)\citenamefont {Egger},
  \citenamefont {Edri}, \citenamefont {Cahen},\ and\ \citenamefont
  {Hodes}}]{Egger2015}%
  \BibitemOpen
  \bibfield  {author} {\bibinfo {author} {\bibfnamefont {David~A.}\
  \bibnamefont {Egger}}, \bibinfo {author} {\bibfnamefont {Eran}\ \bibnamefont
  {Edri}}, \bibinfo {author} {\bibfnamefont {David}\ \bibnamefont {Cahen}}, \
  and\ \bibinfo {author} {\bibfnamefont {Gary}\ \bibnamefont {Hodes}},\
  }\bibfield  {title} {\enquote {\bibinfo {title} {Perovskite solar cells: Do
  we know what we do not know?}}\ }\href {\doibase
  http://dx.doi.org/10.1021/jz502726b} {\bibfield  {journal} {\bibinfo
  {journal} {J. Phys. Chem. Lett.}\ }\textbf {\bibinfo {volume} {6}},\ \bibinfo
  {pages} {279--282} (\bibinfo {year} {2015})}\BibitemShut {NoStop}%
\bibitem [{\citenamefont {McGehee}(2014)}]{hybperov_improvement}%
  \BibitemOpen
  \bibfield  {author} {\bibinfo {author} {\bibfnamefont {Michael~D.}\
  \bibnamefont {McGehee}},\ }\bibfield  {title} {\enquote {\bibinfo {title}
  {Perovskite solar cells: Continuing to soar},}\ }\href {\doibase
  http://dx.doi.org/10.1038/nmat4050} {\bibfield  {journal} {\bibinfo
  {journal} {Nat. Mater.}\ }\textbf {\bibinfo {volume} {13}},\ \bibinfo {pages}
  {845--846} (\bibinfo {year} {2014})}\BibitemShut {NoStop}%
\bibitem [{\citenamefont {Antonietta~Loi}\ and\ \citenamefont
  {Hummelen}(2013)}]{hybperov_cheap}%
  \BibitemOpen
  \bibfield  {author} {\bibinfo {author} {\bibfnamefont {Maria}\ \bibnamefont
  {Antonietta~Loi}}\ and\ \bibinfo {author} {\bibfnamefont {Jan~C.}\
  \bibnamefont {Hummelen}},\ }\bibfield  {title} {\enquote {\bibinfo {title}
  {Hybrid solar cells: Perovskites under the sun},}\ }\href {\doibase
  http://dx.doi.org/10.1038/nmat3815} {\bibfield  {journal} {\bibinfo
  {journal} {Nat. Mater.}\ }\textbf {\bibinfo {volume} {12}},\ \bibinfo {pages}
  {1087--1089} (\bibinfo {year} {2013})}\BibitemShut {NoStop}%
\bibitem [{\citenamefont {Even}\ \emph {et~al.}(2013)\citenamefont {Even},
  \citenamefont {Pedesseau}, \citenamefont {Jancu},\ and\ \citenamefont
  {Katan}}]{hybperov_soc001}%
  \BibitemOpen
  \bibfield  {author} {\bibinfo {author} {\bibfnamefont {Jacky}\ \bibnamefont
  {Even}}, \bibinfo {author} {\bibfnamefont {Laurent}\ \bibnamefont
  {Pedesseau}}, \bibinfo {author} {\bibfnamefont {Jean-Marc}\ \bibnamefont
  {Jancu}}, \ and\ \bibinfo {author} {\bibfnamefont {Claudine}\ \bibnamefont
  {Katan}},\ }\bibfield  {title} {\enquote {\bibinfo {title} {Importance of
  spin–orbit coupling in hybrid organic/inorganic perovskites for
  photovoltaic applications},}\ }\href {\doibase
  http://dx.doi.org/10.1021/jz401532q} {\bibfield  {journal} {\bibinfo
  {journal} {J. Phys. Chem. Lett.}\ }\textbf {\bibinfo {volume} {4}},\ \bibinfo
  {pages} {2999--3005} (\bibinfo {year} {2013})}\BibitemShut {NoStop}%
\bibitem [{\citenamefont {Stroppa}\ \emph {et~al.}(2014)\citenamefont
  {Stroppa}, \citenamefont {Di~Sante}, \citenamefont {Barone}, \citenamefont
  {Bokdam}, \citenamefont {Kresse}, \citenamefont {Franchini}, \citenamefont
  {Whangbo},\ and\ \citenamefont {Picozzi}}]{hybperov_soc002}%
  \BibitemOpen
  \bibfield  {author} {\bibinfo {author} {\bibfnamefont {Alessandro}\
  \bibnamefont {Stroppa}}, \bibinfo {author} {\bibfnamefont {Domenico}\
  \bibnamefont {Di~Sante}}, \bibinfo {author} {\bibfnamefont {Paolo}\
  \bibnamefont {Barone}}, \bibinfo {author} {\bibfnamefont {Menno}\
  \bibnamefont {Bokdam}}, \bibinfo {author} {\bibfnamefont {Georg}\
  \bibnamefont {Kresse}}, \bibinfo {author} {\bibfnamefont {Cesare}\
  \bibnamefont {Franchini}}, \bibinfo {author} {\bibfnamefont {Myung-Hwan}\
  \bibnamefont {Whangbo}}, \ and\ \bibinfo {author} {\bibfnamefont {Silvia}\
  \bibnamefont {Picozzi}},\ }\bibfield  {title} {\enquote {\bibinfo {title}
  {{Tunable ferroelectric polarization and its interplay with
  spin{\textendash}orbit coupling in tin iodide perovskites}},}\ }\href@noop {}
  {\bibfield  {journal} {\bibinfo  {journal} {Nat. Commun.}\ }\textbf {\bibinfo
  {volume} {5}},\ \bibinfo {pages} {5900} (\bibinfo {year} {2014})}\BibitemShut
  {NoStop}%
\bibitem [{\citenamefont {Zhang}\ \emph
  {et~al.}(2015{\natexlab{a}})\citenamefont {Zhang}, \citenamefont {Sun},
  \citenamefont {Sheng}, \citenamefont {Zhai}, \citenamefont {Mielczarek},
  \citenamefont {Zakhidov},\ and\ \citenamefont {Vardeny}}]{hybperov_soc003}%
  \BibitemOpen
  \bibfield  {author} {\bibinfo {author} {\bibfnamefont {C.}~\bibnamefont
  {Zhang}}, \bibinfo {author} {\bibfnamefont {D.}~\bibnamefont {Sun}}, \bibinfo
  {author} {\bibfnamefont {C-X.}\ \bibnamefont {Sheng}}, \bibinfo {author}
  {\bibfnamefont {Y.~X.}\ \bibnamefont {Zhai}}, \bibinfo {author}
  {\bibfnamefont {K.}~\bibnamefont {Mielczarek}}, \bibinfo {author}
  {\bibfnamefont {A.}~\bibnamefont {Zakhidov}}, \ and\ \bibinfo {author}
  {\bibfnamefont {Z.~V.}\ \bibnamefont {Vardeny}},\ }\bibfield  {title}
  {\enquote {\bibinfo {title} {Magnetic field effects in hybrid perovskite
  devices},}\ }\href {\doibase http://dx.doi.org/10.1038/nphys3277} {\bibfield
  {journal} {\bibinfo  {journal} {Nat. Phys.}\ }\textbf {\bibinfo {volume}
  {11}},\ \bibinfo {pages} {427--434} (\bibinfo {year}
  {2015}{\natexlab{a}})}\BibitemShut {NoStop}%
\bibitem [{\citenamefont {Saba}\ \emph {et~al.}(2014)\citenamefont {Saba},
  \citenamefont {Cadelano}, \citenamefont {Marongiu}, \citenamefont {Chen},
  \citenamefont {Sarritzu}, \citenamefont {Sestu}, \citenamefont {Figus},
  \citenamefont {Aresti}, \citenamefont {Piras}, \citenamefont {Geddo~Lehmann},
  \citenamefont {Cannas}, \citenamefont {Musinu}, \citenamefont {Quochi},
  \citenamefont {Mura},\ and\ \citenamefont
  {Bongiovanni}}]{hybperov_excitoncorr}%
  \BibitemOpen
  \bibfield  {author} {\bibinfo {author} {\bibfnamefont {Michele}\ \bibnamefont
  {Saba}}, \bibinfo {author} {\bibfnamefont {Michele}\ \bibnamefont
  {Cadelano}}, \bibinfo {author} {\bibfnamefont {Daniela}\ \bibnamefont
  {Marongiu}}, \bibinfo {author} {\bibfnamefont {Feipeng}\ \bibnamefont
  {Chen}}, \bibinfo {author} {\bibfnamefont {Valerio}\ \bibnamefont
  {Sarritzu}}, \bibinfo {author} {\bibfnamefont {Nicola}\ \bibnamefont
  {Sestu}}, \bibinfo {author} {\bibfnamefont {Cristiana}\ \bibnamefont
  {Figus}}, \bibinfo {author} {\bibfnamefont {Mauro}\ \bibnamefont {Aresti}},
  \bibinfo {author} {\bibfnamefont {Roberto}\ \bibnamefont {Piras}}, \bibinfo
  {author} {\bibfnamefont {Alessandra}\ \bibnamefont {Geddo~Lehmann}}, \bibinfo
  {author} {\bibfnamefont {Carla}\ \bibnamefont {Cannas}}, \bibinfo {author}
  {\bibfnamefont {Anna}\ \bibnamefont {Musinu}}, \bibinfo {author}
  {\bibfnamefont {Francesco}\ \bibnamefont {Quochi}}, \bibinfo {author}
  {\bibfnamefont {Andrea}\ \bibnamefont {Mura}}, \ and\ \bibinfo {author}
  {\bibfnamefont {Giovanni}\ \bibnamefont {Bongiovanni}},\ }\bibfield  {title}
  {\enquote {\bibinfo {title} {Correlated electron--hole plasma in organometal
  perovskites},}\ }\href {\doibase http://dx.doi.org/10.1038/ncomms6049}
  {\bibfield  {journal} {\bibinfo  {journal} {Nat. Commun.}\ }\textbf {\bibinfo
  {volume} {5}},\ \bibinfo {pages} {5049} (\bibinfo {year} {2014})}\BibitemShut
  {NoStop}%
\bibitem [{\citenamefont {Leguy}\ \emph {et~al.}(2015)\citenamefont {Leguy},
  \citenamefont {Frost}, \citenamefont {McMahon}, \citenamefont {Sakai},
  \citenamefont {Kockelmann}, \citenamefont {Law}, \citenamefont {Li},
  \citenamefont {Foglia}, \citenamefont {Walsh}, \citenamefont {O'Regan},
  \citenamefont {Nelson}, \citenamefont {Cabral},\ and\ \citenamefont
  {Barnes}}]{hybperov_dis}%
  \BibitemOpen
  \bibfield  {author} {\bibinfo {author} {\bibfnamefont {Aurelien M.~A.}\
  \bibnamefont {Leguy}}, \bibinfo {author} {\bibfnamefont {Jarvist~Moore}\
  \bibnamefont {Frost}}, \bibinfo {author} {\bibfnamefont {Andrew~P.}\
  \bibnamefont {McMahon}}, \bibinfo {author} {\bibfnamefont {Victoria~Garcia}\
  \bibnamefont {Sakai}}, \bibinfo {author} {\bibfnamefont {W.}~\bibnamefont
  {Kockelmann}}, \bibinfo {author} {\bibfnamefont {ChunHung}\ \bibnamefont
  {Law}}, \bibinfo {author} {\bibfnamefont {Xiaoe}\ \bibnamefont {Li}},
  \bibinfo {author} {\bibfnamefont {Fabrizia}\ \bibnamefont {Foglia}}, \bibinfo
  {author} {\bibfnamefont {Aron}\ \bibnamefont {Walsh}}, \bibinfo {author}
  {\bibfnamefont {Brian~C.}\ \bibnamefont {O'Regan}}, \bibinfo {author}
  {\bibfnamefont {Jenny}\ \bibnamefont {Nelson}}, \bibinfo {author}
  {\bibfnamefont {Joao~T.}\ \bibnamefont {Cabral}}, \ and\ \bibinfo {author}
  {\bibfnamefont {Piers R.~F.}\ \bibnamefont {Barnes}},\ }\bibfield  {title}
  {\enquote {\bibinfo {title} {The dynamics of methylammonium ions in hybrid
  organic-inorganic perovskite solar cells},}\ }\href {\doibase
  http://dx.doi.org/10.1038/ncomms8124} {\bibfield  {journal} {\bibinfo
  {journal} {Nat. Commun.}\ }\textbf {\bibinfo {volume} {6}},\ \bibinfo {pages}
  {7124} (\bibinfo {year} {2015})}\BibitemShut {NoStop}%
\bibitem [{\citenamefont {Motta}\ \emph {et~al.}(2015)\citenamefont {Motta},
  \citenamefont {El-Mellouhi}, \citenamefont {Kais}, \citenamefont {Tabet},
  \citenamefont {Alharbi},\ and\ \citenamefont {Sanvito}}]{hybperov_abinit001}%
  \BibitemOpen
  \bibfield  {author} {\bibinfo {author} {\bibfnamefont {Carlo}\ \bibnamefont
  {Motta}}, \bibinfo {author} {\bibfnamefont {Fedwa}\ \bibnamefont
  {El-Mellouhi}}, \bibinfo {author} {\bibfnamefont {Sabre}\ \bibnamefont
  {Kais}}, \bibinfo {author} {\bibfnamefont {Nouar}\ \bibnamefont {Tabet}},
  \bibinfo {author} {\bibfnamefont {Fahhad}\ \bibnamefont {Alharbi}}, \ and\
  \bibinfo {author} {\bibfnamefont {Stefano}\ \bibnamefont {Sanvito}},\
  }\bibfield  {title} {\enquote {\bibinfo {title} {Revealing the role of
  organic cations in hybrid halide perovskite ch3nh3pbi3},}\ }\href {\doibase
  http://dx.doi.org/10.1038/ncomms8026} {\bibfield  {journal} {\bibinfo
  {journal} {Nat. Commun.}\ }\textbf {\bibinfo {volume} {6}},\ \bibinfo {pages}
  {7026} (\bibinfo {year} {2015})}\BibitemShut {NoStop}%
\bibitem [{\citenamefont {Filip}\ \emph {et~al.}(2014)\citenamefont {Filip},
  \citenamefont {Eperon}, \citenamefont {Snaith},\ and\ \citenamefont
  {Giustino}}]{hybperov_abinit002}%
  \BibitemOpen
  \bibfield  {author} {\bibinfo {author} {\bibfnamefont {Marina~R.}\
  \bibnamefont {Filip}}, \bibinfo {author} {\bibfnamefont {Giles~E.}\
  \bibnamefont {Eperon}}, \bibinfo {author} {\bibfnamefont {Henry~J.}\
  \bibnamefont {Snaith}}, \ and\ \bibinfo {author} {\bibfnamefont {Feliciano}\
  \bibnamefont {Giustino}},\ }\bibfield  {title} {\enquote {\bibinfo {title}
  {Steric engineering of metal-halide perovskites with tunable optical band
  gaps},}\ }\href {\doibase http://dx.doi.org/10.1038/ncomms6757} {\bibfield
  {journal} {\bibinfo  {journal} {Nat. Commun.}\ }\textbf {\bibinfo {volume}
  {5}},\ \bibinfo {pages} {5757} (\bibinfo {year} {2014})}\BibitemShut
  {NoStop}%
\bibitem [{\citenamefont {Saidaminov}\ \emph {et~al.}(2015)\citenamefont
  {Saidaminov}, \citenamefont {Abdelhady}, \citenamefont {Murali},
  \citenamefont {Alarousu}, \citenamefont {Burlakov}, \citenamefont {Peng},
  \citenamefont {Dursun}, \citenamefont {Wang}, \citenamefont {He},
  \citenamefont {Maculan}, \citenamefont {Goriely}, \citenamefont {Wu},
  \citenamefont {Mohammed},\ and\ \citenamefont {Bakr}}]{hybperov_ex001}%
  \BibitemOpen
  \bibfield  {author} {\bibinfo {author} {\bibfnamefont {Makhsud~I.}\
  \bibnamefont {Saidaminov}}, \bibinfo {author} {\bibfnamefont {Ahmed~L.}\
  \bibnamefont {Abdelhady}}, \bibinfo {author} {\bibfnamefont {Banavoth}\
  \bibnamefont {Murali}}, \bibinfo {author} {\bibfnamefont {Erkki}\
  \bibnamefont {Alarousu}}, \bibinfo {author} {\bibfnamefont {Victor~M.}\
  \bibnamefont {Burlakov}}, \bibinfo {author} {\bibfnamefont {Wei}\
  \bibnamefont {Peng}}, \bibinfo {author} {\bibfnamefont {Ibrahim}\
  \bibnamefont {Dursun}}, \bibinfo {author} {\bibfnamefont {Lingfei}\
  \bibnamefont {Wang}}, \bibinfo {author} {\bibfnamefont {Yao}\ \bibnamefont
  {He}}, \bibinfo {author} {\bibfnamefont {Giacomo}\ \bibnamefont {Maculan}},
  \bibinfo {author} {\bibfnamefont {Alain}\ \bibnamefont {Goriely}}, \bibinfo
  {author} {\bibfnamefont {Tom}\ \bibnamefont {Wu}}, \bibinfo {author}
  {\bibfnamefont {Omar~F.}\ \bibnamefont {Mohammed}}, \ and\ \bibinfo {author}
  {\bibfnamefont {Osman~M.}\ \bibnamefont {Bakr}},\ }\bibfield  {title}
  {\enquote {\bibinfo {title} {High-quality bulk hybrid perovskite single
  crystals within minutes by inverse temperature crystallization},}\ }\href
  {\doibase http://dx.doi.org/10.1038/ncomms8586} {\bibfield  {journal}
  {\bibinfo  {journal} {Nat. Commun.}\ }\textbf {\bibinfo {volume} {6}},\
  \bibinfo {pages} {7586} (\bibinfo {year} {2015})}\BibitemShut {NoStop}%
\bibitem [{\citenamefont {Eames}\ \emph {et~al.}(2015)\citenamefont {Eames},
  \citenamefont {Frost}, \citenamefont {Barnes}, \citenamefont {O/'Regan},
  \citenamefont {Walsh},\ and\ \citenamefont {Islam}}]{hybperov_ex002}%
  \BibitemOpen
  \bibfield  {author} {\bibinfo {author} {\bibfnamefont {Christopher}\
  \bibnamefont {Eames}}, \bibinfo {author} {\bibfnamefont {Jarvist~M.}\
  \bibnamefont {Frost}}, \bibinfo {author} {\bibfnamefont {Piers R.~F.}\
  \bibnamefont {Barnes}}, \bibinfo {author} {\bibfnamefont {Brian~C.}\
  \bibnamefont {O/'Regan}}, \bibinfo {author} {\bibfnamefont {Aron}\
  \bibnamefont {Walsh}}, \ and\ \bibinfo {author} {\bibfnamefont {M.~Saiful}\
  \bibnamefont {Islam}},\ }\bibfield  {title} {\enquote {\bibinfo {title}
  {Ionic transport in hybrid lead iodide perovskite solar cells},}\ }\href
  {\doibase http://dx.doi.org/10.1038/ncomms8497} {\bibfield  {journal}
  {\bibinfo  {journal} {Nat. Commun.}\ }\textbf {\bibinfo {volume} {6}},\
  \bibinfo {pages} {7497} (\bibinfo {year} {2015})}\BibitemShut {NoStop}%
\bibitem [{\citenamefont {Heo}\ \emph {et~al.}(2013)\citenamefont {Heo},
  \citenamefont {Im}, \citenamefont {Noh}, \citenamefont {Mandal},
  \citenamefont {Lim}, \citenamefont {Chang}, \citenamefont {Lee},
  \citenamefont {Kim}, \citenamefont {Sarkar}, \citenamefont {K.},
  \citenamefont {Gratzel},\ and\ \citenamefont {Seok}}]{hybperov_ex003}%
  \BibitemOpen
  \bibfield  {author} {\bibinfo {author} {\bibfnamefont {Jin~Hyuck}\
  \bibnamefont {Heo}}, \bibinfo {author} {\bibfnamefont {Sang~Hyuk}\
  \bibnamefont {Im}}, \bibinfo {author} {\bibfnamefont {Jun~Hong}\ \bibnamefont
  {Noh}}, \bibinfo {author} {\bibfnamefont {Tarak~N.}\ \bibnamefont {Mandal}},
  \bibinfo {author} {\bibfnamefont {Choong-Sun}\ \bibnamefont {Lim}}, \bibinfo
  {author} {\bibfnamefont {Jeong~Ah}\ \bibnamefont {Chang}}, \bibinfo {author}
  {\bibfnamefont {Yong~Hui}\ \bibnamefont {Lee}}, \bibinfo {author}
  {\bibfnamefont {Hi-jung}\ \bibnamefont {Kim}}, \bibinfo {author}
  {\bibfnamefont {Arpita}\ \bibnamefont {Sarkar}}, \bibinfo {author}
  {\bibfnamefont {NazeeruddinMd.}\ \bibnamefont {K.}}, \bibinfo {author}
  {\bibfnamefont {Michael}\ \bibnamefont {Gratzel}}, \ and\ \bibinfo {author}
  {\bibfnamefont {Sang~Il}\ \bibnamefont {Seok}},\ }\bibfield  {title}
  {\enquote {\bibinfo {title} {Efficient inorganic-organic hybrid
  heterojunction solar cells containing perovskite compound and polymeric hole
  conductors},}\ }\href {\doibase http://dx.doi.org/10.1038/nphoton.2013.80}
  {\bibfield  {journal} {\bibinfo  {journal} {Nat. Photon.}\ }\textbf {\bibinfo
  {volume} {7}},\ \bibinfo {pages} {486--491} (\bibinfo {year}
  {2013})}\BibitemShut {NoStop}%
\bibitem [{\citenamefont {Young}\ \emph {et~al.}(2012)\citenamefont {Young},
  \citenamefont {Zheng},\ and\ \citenamefont {Rappe}}]{YZR12}%
  \BibitemOpen
  \bibfield  {author} {\bibinfo {author} {\bibfnamefont {Steve~M.}\
  \bibnamefont {Young}}, \bibinfo {author} {\bibfnamefont {Fan}\ \bibnamefont
  {Zheng}}, \ and\ \bibinfo {author} {\bibfnamefont {Andrew~M.}\ \bibnamefont
  {Rappe}},\ }\bibfield  {title} {\enquote {\bibinfo {title} {First-principles
  calculation of the bulk photovoltaic effect in bismuth ferrite},}\
  }\href@noop {} {\bibfield  {journal} {\bibinfo  {journal} {Phys. Rev. Lett.}\
  }\textbf {\bibinfo {volume} {109}},\ \bibinfo {pages} {236601} (\bibinfo
  {year} {2012})}\BibitemShut {NoStop}%
\bibitem [{\citenamefont {{Wang}}\ \emph {et~al.}(2015)\citenamefont {{Wang}},
  \citenamefont {{Young}}, \citenamefont {{Zheng}}, \citenamefont
  {{Grinberg}},\ and\ \citenamefont {{Rappe}}}]{rappe_layeredferro}%
  \BibitemOpen
  \bibfield  {author} {\bibinfo {author} {\bibfnamefont {F.}~\bibnamefont
  {{Wang}}}, \bibinfo {author} {\bibfnamefont {S.~M.}\ \bibnamefont {{Young}}},
  \bibinfo {author} {\bibfnamefont {F.}~\bibnamefont {{Zheng}}}, \bibinfo
  {author} {\bibfnamefont {I.}~\bibnamefont {{Grinberg}}}, \ and\ \bibinfo
  {author} {\bibfnamefont {A.~M.}\ \bibnamefont {{Rappe}}},\ }\bibfield
  {title} {\enquote {\bibinfo {title} {{Bulk photovoltaic effect enhancement
  via electrostatic control in layered ferroelectrics}},}\ }\href@noop {} {\
  (\bibinfo {year} {2015})},\ \Eprint {http://arxiv.org/abs/arXiv:1503.00679}
  {arXiv:1503.00679} \BibitemShut {NoStop}%
\bibitem [{\citenamefont {Wang}\ and\ \citenamefont {Rappe}(2015)}]{Wang2015}%
  \BibitemOpen
  \bibfield  {author} {\bibinfo {author} {\bibfnamefont {Fenggong}\
  \bibnamefont {Wang}}\ and\ \bibinfo {author} {\bibfnamefont {Andrew~M.}\
  \bibnamefont {Rappe}},\ }\bibfield  {title} {\enquote {\bibinfo {title}
  {First-principles calculation of the bulk photovoltaic effect in
  ${\mathrm{knbo}}_{3}$ and
  (k,ba)(ni,nb)${\mathrm{o}}_{3\ensuremath{-}\ensuremath{\delta}}$},}\
  }\href@noop {} {\bibfield  {journal} {\bibinfo  {journal} {Phys. Rev. B}\
  }\textbf {\bibinfo {volume} {91}},\ \bibinfo {pages} {165124} (\bibinfo
  {year} {2015})}\BibitemShut {NoStop}%
\bibitem [{\citenamefont {Mahan}\ and\ \citenamefont {Sofo}(1996)}]{mahan}%
  \BibitemOpen
  \bibfield  {author} {\bibinfo {author} {\bibfnamefont {G.D.}\ \bibnamefont
  {Mahan}}\ and\ \bibinfo {author} {\bibfnamefont {J.O.}\ \bibnamefont
  {Sofo}},\ }\bibfield  {title} {\enquote {\bibinfo {title} {The best
  thermoelectric},}\ }\href@noop {} {\bibfield  {journal} {\bibinfo  {journal}
  {Proc. Natl. Acad. Sci. USA}\ }\textbf {\bibinfo {volume} {93}},\ \bibinfo
  {pages} {7436} (\bibinfo {year} {1996})}\BibitemShut {NoStop}%
\bibitem [{\citenamefont {DiSalvo}(1999)}]{disalvo}%
  \BibitemOpen
  \bibfield  {author} {\bibinfo {author} {\bibfnamefont {F.~J.}\ \bibnamefont
  {DiSalvo}},\ }\bibfield  {title} {\enquote {\bibinfo {title} {Thermoelectric
  cooling and power generation},}\ }\href@noop {} {\bibfield  {journal}
  {\bibinfo  {journal} {Science}\ }\textbf {\bibinfo {volume} {285}},\ \bibinfo
  {pages} {703--706} (\bibinfo {year} {1999})}\BibitemShut {NoStop}%
\bibitem [{\citenamefont {Murphy}\ \emph {et~al.}(2008)\citenamefont {Murphy},
  \citenamefont {Mukerjee},\ and\ \citenamefont {Moore}}]{murphy_moore}%
  \BibitemOpen
  \bibfield  {author} {\bibinfo {author} {\bibfnamefont {Padraig}\ \bibnamefont
  {Murphy}}, \bibinfo {author} {\bibfnamefont {Subroto}\ \bibnamefont
  {Mukerjee}}, \ and\ \bibinfo {author} {\bibfnamefont {Joel}\ \bibnamefont
  {Moore}},\ }\bibfield  {title} {\enquote {\bibinfo {title} {Optimal
  thermoelectric figure of merit of a molecular junction},}\ }\href {\doibase
  10.1103/PhysRevB.78.161406} {\bibfield  {journal} {\bibinfo  {journal} {Phys.
  Rev. B}\ }\textbf {\bibinfo {volume} {78}},\ \bibinfo {pages} {161406}
  (\bibinfo {year} {2008})}\BibitemShut {NoStop}%
\bibitem [{\citenamefont {Van~Hove}(1953)}]{VH53}%
  \BibitemOpen
  \bibfield  {author} {\bibinfo {author} {\bibfnamefont {L\'eon}\ \bibnamefont
  {Van~Hove}},\ }\bibfield  {title} {\enquote {\bibinfo {title} {The occurrence
  of singularities in the elastic frequency distribution of a crystal},}\
  }\href {\doibase 10.1103/PhysRev.89.1189} {\bibfield  {journal} {\bibinfo
  {journal} {Phys. Rev.}\ }\textbf {\bibinfo {volume} {89}},\ \bibinfo {pages}
  {1189--1193} (\bibinfo {year} {1953})}\BibitemShut {NoStop}%
\bibitem [{\citenamefont {Nalwa}(1995)}]{Nalwa1995}%
  \BibitemOpen
  \bibinfo {editor} {\bibfnamefont {Hari~Singh}\ \bibnamefont {Nalwa}},\ ed.,\
  \href@noop {} {\emph {\bibinfo {title} {Ferroelectric Polymers: Chemistry:
  Physics, and Applications}}}\ (\bibinfo  {publisher} {CRC Press},\ \bibinfo
  {year} {1995})\BibitemShut {NoStop}%
\bibitem [{\citenamefont {Lovinger}(1983)}]{L83}%
  \BibitemOpen
  \bibfield  {author} {\bibinfo {author} {\bibfnamefont {Andrew~J}\
  \bibnamefont {Lovinger}},\ }\bibfield  {title} {\enquote {\bibinfo {title}
  {Ferroelectric polymers},}\ }\href@noop {} {\bibfield  {journal} {\bibinfo
  {journal} {Science}\ }\textbf {\bibinfo {volume} {220}},\ \bibinfo {pages}
  {1115--1121} (\bibinfo {year} {1983})}\BibitemShut {NoStop}%
\bibitem [{\citenamefont {Gontia}\ \emph {et~al.}(1999)\citenamefont {Gontia},
  \citenamefont {Frolov}, \citenamefont {Liess}, \citenamefont {Ehrenfreund},
  \citenamefont {Vardeny}, \citenamefont {Tada}, \citenamefont {Kajii},
  \citenamefont {Hidayat}, \citenamefont {Fujii}, \citenamefont {Yoshino},
  \citenamefont {Teraguchi},\ and\ \citenamefont {Masuda}}]{GFL99}%
  \BibitemOpen
  \bibfield  {author} {\bibinfo {author} {\bibfnamefont {I.}~\bibnamefont
  {Gontia}}, \bibinfo {author} {\bibfnamefont {S.~V.}\ \bibnamefont {Frolov}},
  \bibinfo {author} {\bibfnamefont {M.}~\bibnamefont {Liess}}, \bibinfo
  {author} {\bibfnamefont {E.}~\bibnamefont {Ehrenfreund}}, \bibinfo {author}
  {\bibfnamefont {Z.~V.}\ \bibnamefont {Vardeny}}, \bibinfo {author}
  {\bibfnamefont {K.}~\bibnamefont {Tada}}, \bibinfo {author} {\bibfnamefont
  {H.}~\bibnamefont {Kajii}}, \bibinfo {author} {\bibfnamefont
  {R.}~\bibnamefont {Hidayat}}, \bibinfo {author} {\bibfnamefont
  {A.}~\bibnamefont {Fujii}}, \bibinfo {author} {\bibfnamefont
  {K.}~\bibnamefont {Yoshino}}, \bibinfo {author} {\bibfnamefont
  {M.}~\bibnamefont {Teraguchi}}, \ and\ \bibinfo {author} {\bibfnamefont
  {T.}~\bibnamefont {Masuda}},\ }\bibfield  {title} {\enquote {\bibinfo {title}
  {Excitation dynamics in disubstituted polyacetylene},}\ }\href {\doibase
  10.1103/PhysRevLett.82.4058} {\bibfield  {journal} {\bibinfo  {journal}
  {Phys. Rev. Lett.}\ }\textbf {\bibinfo {volume} {82}},\ \bibinfo {pages}
  {4058--4061} (\bibinfo {year} {1999})}\BibitemShut {NoStop}%
\bibitem [{\citenamefont {Rice}\ and\ \citenamefont {Mele}(1982)}]{RM82}%
  \BibitemOpen
  \bibfield  {author} {\bibinfo {author} {\bibfnamefont {M.~J.}\ \bibnamefont
  {Rice}}\ and\ \bibinfo {author} {\bibfnamefont {E.~J.}\ \bibnamefont
  {Mele}},\ }\bibfield  {title} {\enquote {\bibinfo {title} {Elementary
  excitations of a linearly conjugated diatomic polymer},}\ }\href {\doibase
  10.1103/PhysRevLett.49.1455} {\bibfield  {journal} {\bibinfo  {journal}
  {Phys. Rev. Lett.}\ }\textbf {\bibinfo {volume} {49}},\ \bibinfo {pages}
  {1455--1459} (\bibinfo {year} {1982})}\BibitemShut {NoStop}%
\bibitem [{\citenamefont {Geim}\ and\ \citenamefont {Grigorieva}(2013)}]{GG13}%
  \BibitemOpen
  \bibfield  {author} {\bibinfo {author} {\bibfnamefont {AK}~\bibnamefont
  {Geim}}\ and\ \bibinfo {author} {\bibfnamefont {IV}~\bibnamefont
  {Grigorieva}},\ }\bibfield  {title} {\enquote {\bibinfo {title} {Van der
  waals heterostructures},}\ }\href@noop {} {\bibfield  {journal} {\bibinfo
  {journal} {Nature}\ }\textbf {\bibinfo {volume} {499}},\ \bibinfo {pages}
  {419--425} (\bibinfo {year} {2013})}\BibitemShut {NoStop}%
\bibitem [{\citenamefont {Britnell}\ \emph {et~al.}(2013)\citenamefont
  {Britnell}, \citenamefont {Ribeiro}, \citenamefont {Eckmann}, \citenamefont
  {Jalil}, \citenamefont {Belle}, \citenamefont {Mishchenko}, \citenamefont
  {Kim}, \citenamefont {Gorbachev}, \citenamefont {Georgiou}, \citenamefont
  {Morozov}, \citenamefont {Grigorenko}, \citenamefont {Geim}, \citenamefont
  {Casiraghi}, \citenamefont {Neto},\ and\ \citenamefont {Novoselov}}]{BRE13}%
  \BibitemOpen
  \bibfield  {author} {\bibinfo {author} {\bibfnamefont {L.}~\bibnamefont
  {Britnell}}, \bibinfo {author} {\bibfnamefont {R.~M.}\ \bibnamefont
  {Ribeiro}}, \bibinfo {author} {\bibfnamefont {A.}~\bibnamefont {Eckmann}},
  \bibinfo {author} {\bibfnamefont {R.}~\bibnamefont {Jalil}}, \bibinfo
  {author} {\bibfnamefont {B.~D.}\ \bibnamefont {Belle}}, \bibinfo {author}
  {\bibfnamefont {A.}~\bibnamefont {Mishchenko}}, \bibinfo {author}
  {\bibfnamefont {Y.-J.}\ \bibnamefont {Kim}}, \bibinfo {author} {\bibfnamefont
  {R.~V.}\ \bibnamefont {Gorbachev}}, \bibinfo {author} {\bibfnamefont
  {T.}~\bibnamefont {Georgiou}}, \bibinfo {author} {\bibfnamefont {S.~V.}\
  \bibnamefont {Morozov}}, \bibinfo {author} {\bibfnamefont {A.~N.}\
  \bibnamefont {Grigorenko}}, \bibinfo {author} {\bibfnamefont {A.~K.}\
  \bibnamefont {Geim}}, \bibinfo {author} {\bibfnamefont {C.}~\bibnamefont
  {Casiraghi}}, \bibinfo {author} {\bibfnamefont {A.~H.~Castro}\ \bibnamefont
  {Neto}}, \ and\ \bibinfo {author} {\bibfnamefont {K.~S.}\ \bibnamefont
  {Novoselov}},\ }\bibfield  {title} {\enquote {\bibinfo {title} {Strong
  light-matter interactions in heterostructures of atomically thin films},}\
  }\href {\doibase 10.1126/science.1235547} {\bibfield  {journal} {\bibinfo
  {journal} {Science}\ }\textbf {\bibinfo {volume} {340}},\ \bibinfo {pages}
  {1311--1314} (\bibinfo {year} {2013})}\BibitemShut {NoStop}%
\bibitem [{\citenamefont {Yu}\ \emph {et~al.}(2013)\citenamefont {Yu},
  \citenamefont {Liu}, \citenamefont {Zhou}, \citenamefont {Yin}, \citenamefont
  {Li}, \citenamefont {Huang},\ and\ \citenamefont {Duan}}]{YLZ13}%
  \BibitemOpen
  \bibfield  {author} {\bibinfo {author} {\bibfnamefont {Woo~Jong}\
  \bibnamefont {Yu}}, \bibinfo {author} {\bibfnamefont {Yuan}\ \bibnamefont
  {Liu}}, \bibinfo {author} {\bibfnamefont {Hailong}\ \bibnamefont {Zhou}},
  \bibinfo {author} {\bibfnamefont {Anxiang}\ \bibnamefont {Yin}}, \bibinfo
  {author} {\bibfnamefont {Zheng}\ \bibnamefont {Li}}, \bibinfo {author}
  {\bibfnamefont {Yu}~\bibnamefont {Huang}}, \ and\ \bibinfo {author}
  {\bibfnamefont {Xiangfeng}\ \bibnamefont {Duan}},\ }\bibfield  {title}
  {\enquote {\bibinfo {title} {Highly efficient gate-tunable photocurrent
  generation in vertical heterostructures of layered materials},}\ }\href@noop
  {} {\bibfield  {journal} {\bibinfo  {journal} {Nature Nanotech.}\ }\textbf
  {\bibinfo {volume} {8}},\ \bibinfo {pages} {952--958} (\bibinfo {year}
  {2013})}\BibitemShut {NoStop}%
\bibitem [{\citenamefont {Bernardi}\ \emph {et~al.}(2013)\citenamefont
  {Bernardi}, \citenamefont {Palummo},\ and\ \citenamefont {Grossman}}]{BPG13}%
  \BibitemOpen
  \bibfield  {author} {\bibinfo {author} {\bibfnamefont {Marco}\ \bibnamefont
  {Bernardi}}, \bibinfo {author} {\bibfnamefont {Maurizia}\ \bibnamefont
  {Palummo}}, \ and\ \bibinfo {author} {\bibfnamefont {Jeffrey~C}\ \bibnamefont
  {Grossman}},\ }\bibfield  {title} {\enquote {\bibinfo {title} {Extraordinary
  sunlight absorption and one nanometer thick photovoltaics using
  two-dimensional monolayer materials},}\ }\href@noop {} {\bibfield  {journal}
  {\bibinfo  {journal} {Nano lett.}\ }\textbf {\bibinfo {volume} {13}},\
  \bibinfo {pages} {3664--3670} (\bibinfo {year} {2013})}\BibitemShut {NoStop}%
\bibitem [{\citenamefont {Buscema}\ \emph {et~al.}(2014)\citenamefont
  {Buscema}, \citenamefont {Groenendijk}, \citenamefont {Steele}, \citenamefont
  {van~der Zant},\ and\ \citenamefont {Castellanos-Gomez}}]{BGS14}%
  \BibitemOpen
  \bibfield  {author} {\bibinfo {author} {\bibfnamefont {Michele}\ \bibnamefont
  {Buscema}}, \bibinfo {author} {\bibfnamefont {Dirk~J}\ \bibnamefont
  {Groenendijk}}, \bibinfo {author} {\bibfnamefont {Gary~A}\ \bibnamefont
  {Steele}}, \bibinfo {author} {\bibfnamefont {Herre~SJ}\ \bibnamefont {van~der
  Zant}}, \ and\ \bibinfo {author} {\bibfnamefont {Andres}\ \bibnamefont
  {Castellanos-Gomez}},\ }\bibfield  {title} {\enquote {\bibinfo {title}
  {Photovoltaic effect in few-layer black phosphorus pn junctions defined by
  local electrostatic gating},}\ }\href@noop {} {\bibfield  {journal} {\bibinfo
   {journal} {Nature Commun.}\ }\textbf {\bibinfo {volume} {5}} (\bibinfo
  {year} {2014})}\BibitemShut {NoStop}%
\bibitem [{\citenamefont {Singh}\ and\ \citenamefont {Hennig}(2014)}]{SH14}%
  \BibitemOpen
  \bibfield  {author} {\bibinfo {author} {\bibfnamefont {Arunima~K}\
  \bibnamefont {Singh}}\ and\ \bibinfo {author} {\bibfnamefont {Richard~G}\
  \bibnamefont {Hennig}},\ }\bibfield  {title} {\enquote {\bibinfo {title}
  {Computational prediction of two-dimensional group-iv mono-chalcogenides},}\
  }\href@noop {} {\bibfield  {journal} {\bibinfo  {journal} {Appl. Phys.
  Lett.}\ }\textbf {\bibinfo {volume} {105}},\ \bibinfo {pages} {042103}
  (\bibinfo {year} {2014})}\BibitemShut {NoStop}%
\bibitem [{\citenamefont {Gomes}\ and\ \citenamefont {Carvalho}(2015)}]{GC15}%
  \BibitemOpen
  \bibfield  {author} {\bibinfo {author} {\bibfnamefont {Lidia~C}\ \bibnamefont
  {Gomes}}\ and\ \bibinfo {author} {\bibfnamefont {A}~\bibnamefont
  {Carvalho}},\ }\bibfield  {title} {\enquote {\bibinfo {title} {Phosphorene
  analogues: isoelectronic two-dimensional group-iv monochalcogenides with
  orthorhombic structure},}\ }\href@noop {} {\bibfield  {journal} {\bibinfo
  {journal} {arXiv:1504.05627}\ } (\bibinfo {year} {2015})}\BibitemShut
  {NoStop}%
\bibitem [{\citenamefont {Antunez}\ \emph {et~al.}(2011)\citenamefont
  {Antunez}, \citenamefont {Buckley},\ and\ \citenamefont {Brutchey}}]{ABB11}%
  \BibitemOpen
  \bibfield  {author} {\bibinfo {author} {\bibfnamefont {Priscilla~D}\
  \bibnamefont {Antunez}}, \bibinfo {author} {\bibfnamefont {Jannise~J}\
  \bibnamefont {Buckley}}, \ and\ \bibinfo {author} {\bibfnamefont {Richard~L}\
  \bibnamefont {Brutchey}},\ }\bibfield  {title} {\enquote {\bibinfo {title}
  {Tin and germanium monochalcogenide iv--vi semiconductor nanocrystals for use
  in solar cells},}\ }\href@noop {} {\bibfield  {journal} {\bibinfo  {journal}
  {Nanoscale}\ }\textbf {\bibinfo {volume} {3}},\ \bibinfo {pages} {2399--2411}
  (\bibinfo {year} {2011})}\BibitemShut {NoStop}%
\bibitem [{\citenamefont {Li}\ \emph {et~al.}(2016)\citenamefont {Li},
  \citenamefont {Liu}, \citenamefont {Wang},\ and\ \citenamefont {Li}}]{LLW16}%
  \BibitemOpen
  \bibfield  {author} {\bibinfo {author} {\bibfnamefont {Feng}\ \bibnamefont
  {Li}}, \bibinfo {author} {\bibfnamefont {Xiuhong}\ \bibnamefont {Liu}},
  \bibinfo {author} {\bibfnamefont {Yu}~\bibnamefont {Wang}}, \ and\ \bibinfo
  {author} {\bibfnamefont {Yafei}\ \bibnamefont {Li}},\ }\bibfield  {title}
  {\enquote {\bibinfo {title} {Germanium monosulfide monolayer: a novel
  two-dimensional semiconductor with a high carrier mobility},}\ }\href@noop {}
  {\bibfield  {journal} {\bibinfo  {journal} {J. Mater. Chem. C}\ }\textbf
  {\bibinfo {volume} {4}},\ \bibinfo {pages} {2155--2159} (\bibinfo {year}
  {2016})}\BibitemShut {NoStop}%
\bibitem [{\citenamefont {Rodin}\ \emph {et~al.}(2016)\citenamefont {Rodin},
  \citenamefont {Gomes}, \citenamefont {Carvalho},\ and\ \citenamefont
  {Castro~Neto}}]{RGC16}%
  \BibitemOpen
  \bibfield  {author} {\bibinfo {author} {\bibfnamefont {A.~S.}\ \bibnamefont
  {Rodin}}, \bibinfo {author} {\bibfnamefont {Lidia~C.}\ \bibnamefont {Gomes}},
  \bibinfo {author} {\bibfnamefont {A.}~\bibnamefont {Carvalho}}, \ and\
  \bibinfo {author} {\bibfnamefont {A.~H.}\ \bibnamefont {Castro~Neto}},\
  }\bibfield  {title} {\enquote {\bibinfo {title} {Valley physics in tin (ii)
  sulfide},}\ }\href {\doibase 10.1103/PhysRevB.93.045431} {\bibfield
  {journal} {\bibinfo  {journal} {Phys. Rev. B}\ }\textbf {\bibinfo {volume}
  {93}},\ \bibinfo {pages} {045431} (\bibinfo {year} {2016})}\BibitemShut
  {NoStop}%
\bibitem [{\citenamefont {Li}\ \emph {et~al.}(2012)\citenamefont {Li},
  \citenamefont {Huang}, \citenamefont {Snigdha}, \citenamefont {Yu},\ and\
  \citenamefont {Cao}}]{LHS12}%
  \BibitemOpen
  \bibfield  {author} {\bibinfo {author} {\bibfnamefont {Chun}\ \bibnamefont
  {Li}}, \bibinfo {author} {\bibfnamefont {Liang}\ \bibnamefont {Huang}},
  \bibinfo {author} {\bibfnamefont {Gayatri~Pongur}\ \bibnamefont {Snigdha}},
  \bibinfo {author} {\bibfnamefont {Yifei}\ \bibnamefont {Yu}}, \ and\ \bibinfo
  {author} {\bibfnamefont {Linyou}\ \bibnamefont {Cao}},\ }\bibfield  {title}
  {\enquote {\bibinfo {title} {Role of boundary layer diffusion in vapor
  deposition growth of chalcogenide nanosheets: The case of ges},}\ }\href@noop
  {} {\bibfield  {journal} {\bibinfo  {journal} {ACS Nano}\ }\textbf {\bibinfo
  {volume} {6}},\ \bibinfo {pages} {8868--8877} (\bibinfo {year}
  {2012})}\BibitemShut {NoStop}%
\bibitem [{\citenamefont {Ulaganathan}\ \emph {et~al.}(2016)\citenamefont
  {Ulaganathan}, \citenamefont {Lu}, \citenamefont {Kuo}, \citenamefont
  {Tamalampudi}, \citenamefont {Sankar}, \citenamefont {Boopathi},
  \citenamefont {Anand}, \citenamefont {Yadav}, \citenamefont {Mathew},
  \citenamefont {Liu} \emph {et~al.}}]{ULK16}%
  \BibitemOpen
  \bibfield  {author} {\bibinfo {author} {\bibfnamefont {Rajesh~Kumar}\
  \bibnamefont {Ulaganathan}}, \bibinfo {author} {\bibfnamefont {Yi-Ying}\
  \bibnamefont {Lu}}, \bibinfo {author} {\bibfnamefont {Chia-Jung}\
  \bibnamefont {Kuo}}, \bibinfo {author} {\bibfnamefont {Srinivasa~Reddy}\
  \bibnamefont {Tamalampudi}}, \bibinfo {author} {\bibfnamefont {Raman}\
  \bibnamefont {Sankar}}, \bibinfo {author} {\bibfnamefont
  {Karunakara~Moorthy}\ \bibnamefont {Boopathi}}, \bibinfo {author}
  {\bibfnamefont {Ankur}\ \bibnamefont {Anand}}, \bibinfo {author}
  {\bibfnamefont {Kanchan}\ \bibnamefont {Yadav}}, \bibinfo {author}
  {\bibfnamefont {Roshan~Jesus}\ \bibnamefont {Mathew}}, \bibinfo {author}
  {\bibfnamefont {Chia-Rung}\ \bibnamefont {Liu}},  \emph {et~al.},\ }\bibfield
   {title} {\enquote {\bibinfo {title} {High photosensitivity and broad
  spectral response of multi-layered germanium sulfide transistors},}\
  }\href@noop {} {\bibfield  {journal} {\bibinfo  {journal} {Nanoscale}\
  }\textbf {\bibinfo {volume} {8}},\ \bibinfo {pages} {2284--2292} (\bibinfo
  {year} {2016})}\BibitemShut {NoStop}%
\bibitem [{\citenamefont {Vaughn~II}\ \emph {et~al.}(2010)\citenamefont
  {Vaughn~II}, \citenamefont {Patel}, \citenamefont {Hickner},\ and\
  \citenamefont {Schaak}}]{VPH10}%
  \BibitemOpen
  \bibfield  {author} {\bibinfo {author} {\bibfnamefont {Dimitri~D}\
  \bibnamefont {Vaughn~II}}, \bibinfo {author} {\bibfnamefont {Romesh~J}\
  \bibnamefont {Patel}}, \bibinfo {author} {\bibfnamefont {Michael~A}\
  \bibnamefont {Hickner}}, \ and\ \bibinfo {author} {\bibfnamefont {Raymond~E}\
  \bibnamefont {Schaak}},\ }\bibfield  {title} {\enquote {\bibinfo {title}
  {Single-crystal colloidal nanosheets of ges and gese},}\ }\href@noop {}
  {\bibfield  {journal} {\bibinfo  {journal} {J. Amer. Chem. Soc.}\ }\textbf
  {\bibinfo {volume} {132}},\ \bibinfo {pages} {15170--15172} (\bibinfo {year}
  {2010})}\BibitemShut {NoStop}%
\bibitem [{\citenamefont {Ramasamy}\ \emph {et~al.}(2016)\citenamefont
  {Ramasamy}, \citenamefont {Kwak}, \citenamefont {Lim}, \citenamefont {Ra},\
  and\ \citenamefont {Lee}}]{RKD16}%
  \BibitemOpen
  \bibfield  {author} {\bibinfo {author} {\bibfnamefont {Parthiban}\
  \bibnamefont {Ramasamy}}, \bibinfo {author} {\bibfnamefont {Dohyun}\
  \bibnamefont {Kwak}}, \bibinfo {author} {\bibfnamefont {Da-Hye}\ \bibnamefont
  {Lim}}, \bibinfo {author} {\bibfnamefont {Hyun-Soo}\ \bibnamefont {Ra}}, \
  and\ \bibinfo {author} {\bibfnamefont {Jong-Soo}\ \bibnamefont {Lee}},\
  }\bibfield  {title} {\enquote {\bibinfo {title} {Solution synthesis of ges
  and gese nanosheets for high-sensitivity photodetectors},}\ }\href@noop {}
  {\bibfield  {journal} {\bibinfo  {journal} {J. Mater. Chem. C}\ }\textbf
  {\bibinfo {volume} {4}},\ \bibinfo {pages} {479--485} (\bibinfo {year}
  {2016})}\BibitemShut {NoStop}%
\bibitem [{\citenamefont {Brent}\ \emph {et~al.}(2015)\citenamefont {Brent},
  \citenamefont {Lewis}, \citenamefont {Lorenz}, \citenamefont {Lewis},
  \citenamefont {Savjani}, \citenamefont {Haigh}, \citenamefont {Seifert},
  \citenamefont {Derby},\ and\ \citenamefont {O’Brien}}]{BLL15}%
  \BibitemOpen
  \bibfield  {author} {\bibinfo {author} {\bibfnamefont {Jack~R}\ \bibnamefont
  {Brent}}, \bibinfo {author} {\bibfnamefont {David~J}\ \bibnamefont {Lewis}},
  \bibinfo {author} {\bibfnamefont {Tommy}\ \bibnamefont {Lorenz}}, \bibinfo
  {author} {\bibfnamefont {Edward~A}\ \bibnamefont {Lewis}}, \bibinfo {author}
  {\bibfnamefont {Nicky}\ \bibnamefont {Savjani}}, \bibinfo {author}
  {\bibfnamefont {Sarah~J}\ \bibnamefont {Haigh}}, \bibinfo {author}
  {\bibfnamefont {Gotthard}\ \bibnamefont {Seifert}}, \bibinfo {author}
  {\bibfnamefont {Brian}\ \bibnamefont {Derby}}, \ and\ \bibinfo {author}
  {\bibfnamefont {Paul}\ \bibnamefont {O’Brien}},\ }\bibfield  {title}
  {\enquote {\bibinfo {title} {Tin (ii) sulfide (sns) nanosheets by
  liquid-phase exfoliation of herzenbergite: Iv--vi main group two-dimensional
  atomic crystals},}\ }\href@noop {} {\bibfield  {journal} {\bibinfo  {journal}
  {J. Amer. Chem. Soc.}\ }\textbf {\bibinfo {volume} {137}},\ \bibinfo {pages}
  {12689--12696} (\bibinfo {year} {2015})}\BibitemShut {NoStop}%
\bibitem [{\citenamefont {Xia}\ \emph {et~al.}(2016)\citenamefont {Xia},
  \citenamefont {Li}, \citenamefont {Huang}, \citenamefont {Mao}, \citenamefont
  {Zhu}, \citenamefont {Wang}, \citenamefont {Xu},\ and\ \citenamefont
  {Meng}}]{XLH16}%
  \BibitemOpen
  \bibfield  {author} {\bibinfo {author} {\bibfnamefont {Jing}\ \bibnamefont
  {Xia}}, \bibinfo {author} {\bibfnamefont {Xuan-Ze}\ \bibnamefont {Li}},
  \bibinfo {author} {\bibfnamefont {Xing}\ \bibnamefont {Huang}}, \bibinfo
  {author} {\bibfnamefont {Nannan}\ \bibnamefont {Mao}}, \bibinfo {author}
  {\bibfnamefont {Dan-Dan}\ \bibnamefont {Zhu}}, \bibinfo {author}
  {\bibfnamefont {Lei}\ \bibnamefont {Wang}}, \bibinfo {author} {\bibfnamefont
  {Hua}\ \bibnamefont {Xu}}, \ and\ \bibinfo {author} {\bibfnamefont
  {Xiang-Min}\ \bibnamefont {Meng}},\ }\bibfield  {title} {\enquote {\bibinfo
  {title} {Physical vapor deposition synthesis of two-dimensional orthorhombic
  sns flakes with strong angle/temperature-dependent raman responses},}\
  }\href@noop {} {\bibfield  {journal} {\bibinfo  {journal} {Nanoscale}\
  }\textbf {\bibinfo {volume} {8}},\ \bibinfo {pages} {2063--2070} (\bibinfo
  {year} {2016})}\BibitemShut {NoStop}%
\bibitem [{\citenamefont {Li}\ \emph {et~al.}(2013)\citenamefont {Li},
  \citenamefont {Chen}, \citenamefont {Hu}, \citenamefont {Wang}, \citenamefont
  {Zhang}, \citenamefont {Chen},\ and\ \citenamefont {Wang}}]{LCH13}%
  \BibitemOpen
  \bibfield  {author} {\bibinfo {author} {\bibfnamefont {Lun}\ \bibnamefont
  {Li}}, \bibinfo {author} {\bibfnamefont {Zhong}\ \bibnamefont {Chen}},
  \bibinfo {author} {\bibfnamefont {Ying}\ \bibnamefont {Hu}}, \bibinfo
  {author} {\bibfnamefont {Xuewen}\ \bibnamefont {Wang}}, \bibinfo {author}
  {\bibfnamefont {Ting}\ \bibnamefont {Zhang}}, \bibinfo {author}
  {\bibfnamefont {Wei}\ \bibnamefont {Chen}}, \ and\ \bibinfo {author}
  {\bibfnamefont {Qiangbin}\ \bibnamefont {Wang}},\ }\bibfield  {title}
  {\enquote {\bibinfo {title} {Single-layer single-crystalline snse
  nanosheets},}\ }\href {\doibase 10.1021/ja3108017} {\bibfield  {journal}
  {\bibinfo  {journal} {J. Am. Chem. Soc.}\ }\textbf {\bibinfo {volume}
  {135}},\ \bibinfo {pages} {1213--1216} (\bibinfo {year} {2013})}\BibitemShut
  {NoStop}%
\bibitem [{\citenamefont {Zhang}\ \emph
  {et~al.}(2015{\natexlab{b}})\citenamefont {Zhang}, \citenamefont {Zhu},
  \citenamefont {Wu}, \citenamefont {Cui}, \citenamefont {Li}, \citenamefont
  {Jiang}, \citenamefont {Gao}, \citenamefont {Wang},\ and\ \citenamefont
  {Cui}}]{ZZW15}%
  \BibitemOpen
  \bibfield  {author} {\bibinfo {author} {\bibfnamefont {Jian}\ \bibnamefont
  {Zhang}}, \bibinfo {author} {\bibfnamefont {Hongyang}\ \bibnamefont {Zhu}},
  \bibinfo {author} {\bibfnamefont {Xiaoxin}\ \bibnamefont {Wu}}, \bibinfo
  {author} {\bibfnamefont {Hang}\ \bibnamefont {Cui}}, \bibinfo {author}
  {\bibfnamefont {Dongmei}\ \bibnamefont {Li}}, \bibinfo {author}
  {\bibfnamefont {Junru}\ \bibnamefont {Jiang}}, \bibinfo {author}
  {\bibfnamefont {Chunxiao}\ \bibnamefont {Gao}}, \bibinfo {author}
  {\bibfnamefont {Qiushi}\ \bibnamefont {Wang}}, \ and\ \bibinfo {author}
  {\bibfnamefont {Qiliang}\ \bibnamefont {Cui}},\ }\bibfield  {title} {\enquote
  {\bibinfo {title} {Plasma-assisted synthesis and pressure-induced structural
  transition of single-crystalline snse nanosheets},}\ }\href@noop {}
  {\bibfield  {journal} {\bibinfo  {journal} {Nanoscale}\ }\textbf {\bibinfo
  {volume} {7}},\ \bibinfo {pages} {10807--10816} (\bibinfo {year}
  {2015}{\natexlab{b}})}\BibitemShut {NoStop}%
\bibitem [{\citenamefont {Zhao}\ \emph {et~al.}(2015)\citenamefont {Zhao},
  \citenamefont {Wang}, \citenamefont {Zhou}, \citenamefont {Liao},
  \citenamefont {Jiang}, \citenamefont {Yang}, \citenamefont {Chen},
  \citenamefont {Lin}, \citenamefont {Wang}, \citenamefont {Peng} \emph
  {et~al.}}]{ZWZ15}%
  \BibitemOpen
  \bibfield  {author} {\bibinfo {author} {\bibfnamefont {Shuli}\ \bibnamefont
  {Zhao}}, \bibinfo {author} {\bibfnamefont {Huan}\ \bibnamefont {Wang}},
  \bibinfo {author} {\bibfnamefont {Yu}~\bibnamefont {Zhou}}, \bibinfo {author}
  {\bibfnamefont {Lei}\ \bibnamefont {Liao}}, \bibinfo {author} {\bibfnamefont
  {Ying}\ \bibnamefont {Jiang}}, \bibinfo {author} {\bibfnamefont {Xiao}\
  \bibnamefont {Yang}}, \bibinfo {author} {\bibfnamefont {Guanchu}\
  \bibnamefont {Chen}}, \bibinfo {author} {\bibfnamefont {Min}\ \bibnamefont
  {Lin}}, \bibinfo {author} {\bibfnamefont {Yong}\ \bibnamefont {Wang}},
  \bibinfo {author} {\bibfnamefont {Hailin}\ \bibnamefont {Peng}},  \emph
  {et~al.},\ }\bibfield  {title} {\enquote {\bibinfo {title} {Controlled
  synthesis of single-crystal snse nanoplates},}\ }\href@noop {} {\bibfield
  {journal} {\bibinfo  {journal} {Nano Research}\ }\textbf {\bibinfo {volume}
  {8}},\ \bibinfo {pages} {288--295} (\bibinfo {year} {2015})}\BibitemShut
  {NoStop}%
\bibitem [{\citenamefont {Bena}\ and\ \citenamefont {Montambaux}(2009)}]{BM09}%
  \BibitemOpen
  \bibfield  {author} {\bibinfo {author} {\bibfnamefont {Cristina}\
  \bibnamefont {Bena}}\ and\ \bibinfo {author} {\bibfnamefont {Gilles}\
  \bibnamefont {Montambaux}},\ }\bibfield  {title} {\enquote {\bibinfo {title}
  {Remarks on the tight-binding model of graphene},}\ }\href@noop {} {\bibfield
   {journal} {\bibinfo  {journal} {New J. Phys.}\ }\textbf {\bibinfo {volume}
  {11}},\ \bibinfo {pages} {095003} (\bibinfo {year} {2009})}\BibitemShut
  {NoStop}%
\bibitem [{\citenamefont {Dobard{\v{z}}i{\'c}}\ \emph
  {et~al.}(2015)\citenamefont {Dobard{\v{z}}i{\'c}}, \citenamefont
  {Dimitrijevi{\'c}},\ and\ \citenamefont {Milovanovi{\'c}}}]{DDM15}%
  \BibitemOpen
  \bibfield  {author} {\bibinfo {author} {\bibfnamefont {E}~\bibnamefont
  {Dobard{\v{z}}i{\'c}}}, \bibinfo {author} {\bibfnamefont {M}~\bibnamefont
  {Dimitrijevi{\'c}}}, \ and\ \bibinfo {author} {\bibfnamefont
  {MV}~\bibnamefont {Milovanovi{\'c}}},\ }\bibfield  {title} {\enquote
  {\bibinfo {title} {Generalized bloch theorem and topological
  characterization},}\ }\href {\doibase 10.1103/PhysRevB.91.125424} {\bibfield
  {journal} {\bibinfo  {journal} {Phys. Rev. B}\ }\textbf {\bibinfo {volume}
  {91}},\ \bibinfo {pages} {125424} (\bibinfo {year} {2015})}\BibitemShut
  {NoStop}%
\bibitem [{\citenamefont {Fruchart}\ \emph {et~al.}(2014)\citenamefont
  {Fruchart}, \citenamefont {Carpentier},\ and\ \citenamefont
  {Gawedzki}}]{FCG14}%
  \BibitemOpen
  \bibfield  {author} {\bibinfo {author} {\bibfnamefont {Michel}\ \bibnamefont
  {Fruchart}}, \bibinfo {author} {\bibfnamefont {David}\ \bibnamefont
  {Carpentier}}, \ and\ \bibinfo {author} {\bibfnamefont {Krzysztof}\
  \bibnamefont {Gawedzki}},\ }\bibfield  {title} {\enquote {\bibinfo {title}
  {Parallel transport and band theory in crystals},}\ }\href@noop {} {\bibfield
   {journal} {\bibinfo  {journal} {Europhys. Lett.}\ }\textbf {\bibinfo
  {volume} {106}},\ \bibinfo {pages} {60002} (\bibinfo {year}
  {2014})}\BibitemShut {NoStop}%
\bibitem [{\citenamefont {Banerjee}\ \emph {et~al.}(2009)\citenamefont
  {Banerjee}, \citenamefont {Singh}, \citenamefont {Pardo},\ and\ \citenamefont
  {Pickett}}]{BSP09}%
  \BibitemOpen
  \bibfield  {author} {\bibinfo {author} {\bibfnamefont {S}~\bibnamefont
  {Banerjee}}, \bibinfo {author} {\bibfnamefont {RRP}\ \bibnamefont {Singh}},
  \bibinfo {author} {\bibfnamefont {V}~\bibnamefont {Pardo}}, \ and\ \bibinfo
  {author} {\bibfnamefont {WE}~\bibnamefont {Pickett}},\ }\bibfield  {title}
  {\enquote {\bibinfo {title} {Tight-binding modeling and low-energy behavior
  of the semi-dirac point},}\ }\href@noop {} {\bibfield  {journal} {\bibinfo
  {journal} {Phys. Rev. Lett.}\ }\textbf {\bibinfo {volume} {103}},\ \bibinfo
  {pages} {016402} (\bibinfo {year} {2009})}\BibitemShut {NoStop}%
\bibitem [{\citenamefont {Su}\ \emph {et~al.}(1979)\citenamefont {Su},
  \citenamefont {Schrieffer},\ and\ \citenamefont {Heeger}}]{SSH79}%
  \BibitemOpen
  \bibfield  {author} {\bibinfo {author} {\bibfnamefont {W.~P.}\ \bibnamefont
  {Su}}, \bibinfo {author} {\bibfnamefont {J.~R.}\ \bibnamefont {Schrieffer}},
  \ and\ \bibinfo {author} {\bibfnamefont {A.~J.}\ \bibnamefont {Heeger}},\
  }\bibfield  {title} {\enquote {\bibinfo {title} {Solitons in
  polyacetylene},}\ }\href {\doibase 10.1103/PhysRevLett.42.1698} {\bibfield
  {journal} {\bibinfo  {journal} {Phys. Rev. Lett.}\ }\textbf {\bibinfo
  {volume} {42}},\ \bibinfo {pages} {1698--1701} (\bibinfo {year}
  {1979})}\BibitemShut {NoStop}%
\bibitem [{\citenamefont {Xia}\ \emph {et~al.}(2014)\citenamefont {Xia},
  \citenamefont {Wang},\ and\ \citenamefont {Jia}}]{XWJ14}%
  \BibitemOpen
  \bibfield  {author} {\bibinfo {author} {\bibfnamefont {Fengnian}\
  \bibnamefont {Xia}}, \bibinfo {author} {\bibfnamefont {Han}\ \bibnamefont
  {Wang}}, \ and\ \bibinfo {author} {\bibfnamefont {Yichen}\ \bibnamefont
  {Jia}},\ }\bibfield  {title} {\enquote {\bibinfo {title} {Rediscovering black
  phosphorus as an anisotropic layered material for optoelectronics and
  electronics},}\ }\href@noop {} {\bibfield  {journal} {\bibinfo  {journal}
  {Nature Commun.}\ }\textbf {\bibinfo {volume} {5}} (\bibinfo {year}
  {2014})}\BibitemShut {NoStop}%
\bibitem [{\citenamefont {Liu}\ \emph {et~al.}(2015)\citenamefont {Liu},
  \citenamefont {Fu}, \citenamefont {Wang}, \citenamefont {Feng}, \citenamefont
  {Liu}, \citenamefont {Wan}, \citenamefont {Zhou}, \citenamefont {Wang},
  \citenamefont {Shao}, \citenamefont {Ho} \emph {et~al.}}]{LFW15}%
  \BibitemOpen
  \bibfield  {author} {\bibinfo {author} {\bibfnamefont {Erfu}\ \bibnamefont
  {Liu}}, \bibinfo {author} {\bibfnamefont {Yajun}\ \bibnamefont {Fu}},
  \bibinfo {author} {\bibfnamefont {Yaojia}\ \bibnamefont {Wang}}, \bibinfo
  {author} {\bibfnamefont {Yanqing}\ \bibnamefont {Feng}}, \bibinfo {author}
  {\bibfnamefont {Huimei}\ \bibnamefont {Liu}}, \bibinfo {author}
  {\bibfnamefont {Xiangang}\ \bibnamefont {Wan}}, \bibinfo {author}
  {\bibfnamefont {Wei}\ \bibnamefont {Zhou}}, \bibinfo {author} {\bibfnamefont
  {Baigeng}\ \bibnamefont {Wang}}, \bibinfo {author} {\bibfnamefont {Lubin}\
  \bibnamefont {Shao}}, \bibinfo {author} {\bibfnamefont {Ching-Hwa}\
  \bibnamefont {Ho}},  \emph {et~al.},\ }\bibfield  {title} {\enquote {\bibinfo
  {title} {Integrated digital inverters based on two-dimensional anisotropic
  res2 field-effect transistors},}\ }\href@noop {} {\bibfield  {journal}
  {\bibinfo  {journal} {Nature Commun.}\ }\textbf {\bibinfo {volume} {6}}
  (\bibinfo {year} {2015})}\BibitemShut {NoStop}%
\bibitem [{\citenamefont {Rudenko}\ and\ \citenamefont
  {Katsnelson}(2014)}]{RK14}%
  \BibitemOpen
  \bibfield  {author} {\bibinfo {author} {\bibfnamefont {A.~N.}\ \bibnamefont
  {Rudenko}}\ and\ \bibinfo {author} {\bibfnamefont {M.~I.}\ \bibnamefont
  {Katsnelson}},\ }\bibfield  {title} {\enquote {\bibinfo {title}
  {Quasiparticle band structure and tight-binding model for single- and bilayer
  black phosphorus},}\ }\href {\doibase 10.1103/PhysRevB.89.201408} {\bibfield
  {journal} {\bibinfo  {journal} {Phys. Rev. B}\ }\textbf {\bibinfo {volume}
  {89}},\ \bibinfo {pages} {201408} (\bibinfo {year} {2014})}\BibitemShut
  {NoStop}%
\bibitem [{\citenamefont {Rudenko}\ \emph {et~al.}(2015)\citenamefont
  {Rudenko}, \citenamefont {Yuan},\ and\ \citenamefont {Katsnelson}}]{RYK15}%
  \BibitemOpen
  \bibfield  {author} {\bibinfo {author} {\bibfnamefont {AN}~\bibnamefont
  {Rudenko}}, \bibinfo {author} {\bibfnamefont {Shengjun}\ \bibnamefont
  {Yuan}}, \ and\ \bibinfo {author} {\bibfnamefont {MI}~\bibnamefont
  {Katsnelson}},\ }\bibfield  {title} {\enquote {\bibinfo {title} {Toward a
  realistic description of multilayer black phosphorus: from $ gw $
  approximation to large-scale tight-binding simulations},}\ }\href@noop {}
  {\bibfield  {journal} {\bibinfo  {journal} {arXiv:1506.01954}\ } (\bibinfo
  {year} {2015})}\BibitemShut {NoStop}%
\bibitem [{\citenamefont {Ezawa}(2014)}]{E14}%
  \BibitemOpen
  \bibfield  {author} {\bibinfo {author} {\bibfnamefont {Motohiko}\
  \bibnamefont {Ezawa}},\ }\bibfield  {title} {\enquote {\bibinfo {title}
  {Topological origin of quasi-flat edge band in phosphorene},}\ }\href
  {http://stacks.iop.org/1367-2630/16/i=11/a=115004} {\bibfield  {journal}
  {\bibinfo  {journal} {New J. Phys.}\ }\textbf {\bibinfo {volume} {16}},\
  \bibinfo {pages} {115004} (\bibinfo {year} {2014})}\BibitemShut {NoStop}%
\bibitem [{\citenamefont {Montambaux}\ \emph {et~al.}(2009)\citenamefont
  {Montambaux}, \citenamefont {Pi\'echon}, \citenamefont {Fuchs},\ and\
  \citenamefont {Goerbig}}]{MPF09}%
  \BibitemOpen
  \bibfield  {author} {\bibinfo {author} {\bibfnamefont {G.}~\bibnamefont
  {Montambaux}}, \bibinfo {author} {\bibfnamefont {F.}~\bibnamefont
  {Pi\'echon}}, \bibinfo {author} {\bibfnamefont {J.-N.}\ \bibnamefont
  {Fuchs}}, \ and\ \bibinfo {author} {\bibfnamefont {M.~O.}\ \bibnamefont
  {Goerbig}},\ }\bibfield  {title} {\enquote {\bibinfo {title} {Merging of
  dirac points in a two-dimensional crystal},}\ }\href {\doibase
  10.1103/PhysRevB.80.153412} {\bibfield  {journal} {\bibinfo  {journal} {Phys.
  Rev. B}\ }\textbf {\bibinfo {volume} {80}},\ \bibinfo {pages} {153412}
  (\bibinfo {year} {2009})}\BibitemShut {NoStop}%
\bibitem [{\citenamefont {Pagliaro}\ \emph {et~al.}(2008)\citenamefont
  {Pagliaro}, \citenamefont {Palmisano},\ and\ \citenamefont
  {Ciriminna}}]{flexible}%
  \BibitemOpen
  \bibfield  {author} {\bibinfo {author} {\bibfnamefont {Mario}\ \bibnamefont
  {Pagliaro}}, \bibinfo {author} {\bibfnamefont {Giovanni}\ \bibnamefont
  {Palmisano}}, \ and\ \bibinfo {author} {\bibfnamefont {Rosaria}\ \bibnamefont
  {Ciriminna}},\ }\href@noop {} {\emph {\bibinfo {title} {Flexible solar
  cells}}}\ (\bibinfo  {publisher} {Wiley},\ \bibinfo {year}
  {2008})\BibitemShut {NoStop}%
\bibitem [{\citenamefont {Perdew}\ \emph {et~al.}(1996)\citenamefont {Perdew},
  \citenamefont {Burke},\ and\ \citenamefont {Ernzerhof}}]{pbe1997}%
  \BibitemOpen
  \bibfield  {author} {\bibinfo {author} {\bibfnamefont {John~P.}\ \bibnamefont
  {Perdew}}, \bibinfo {author} {\bibfnamefont {Kieron}\ \bibnamefont {Burke}},
  \ and\ \bibinfo {author} {\bibfnamefont {Matthias}\ \bibnamefont
  {Ernzerhof}},\ }\bibfield  {title} {\enquote {\bibinfo {title} {Generalized
  gradient approximation made simple},}\ }\href {\doibase
  10.1103/PhysRevLett.77.3865} {\bibfield  {journal} {\bibinfo  {journal}
  {Phys. Rev. Lett.}\ }\textbf {\bibinfo {volume} {77}},\ \bibinfo {pages}
  {3865--3868} (\bibinfo {year} {1996})}\BibitemShut {NoStop}%
\bibitem [{\citenamefont {Garrity}\ \emph {et~al.}(2014)\citenamefont
  {Garrity}, \citenamefont {Bennett}, \citenamefont {Rabe},\ and\ \citenamefont
  {Vanderbilt}}]{Garrity2014}%
  \BibitemOpen
  \bibfield  {author} {\bibinfo {author} {\bibfnamefont {Kevin~F.}\
  \bibnamefont {Garrity}}, \bibinfo {author} {\bibfnamefont {Joseph~W.}\
  \bibnamefont {Bennett}}, \bibinfo {author} {\bibfnamefont {Karin~M.}\
  \bibnamefont {Rabe}}, \ and\ \bibinfo {author} {\bibfnamefont {David}\
  \bibnamefont {Vanderbilt}},\ }\bibfield  {title} {\enquote {\bibinfo {title}
  {Pseudopotentials for high-throughput \{DFT\} calculations},}\ }\href
  {\doibase http://dx.doi.org/10.1016/j.commatsci.2013.08.053} {\bibfield
  {journal} {\bibinfo  {journal} {Comput. Mater. Sci.}\ }\textbf {\bibinfo
  {volume} {81}},\ \bibinfo {pages} {446 -- 452} (\bibinfo {year}
  {2014})}\BibitemShut {NoStop}%
\bibitem [{\citenamefont {Giannozzi}\ \emph {et~al.}(2009)\citenamefont
  {Giannozzi} \emph {et~al.}}]{giannozzi}%
  \BibitemOpen
  \bibfield  {author} {\bibinfo {author} {\bibfnamefont {Paolo}\ \bibnamefont
  {Giannozzi}} \emph {et~al.},\ }\bibfield  {title} {\enquote {\bibinfo {title}
  {Quantum espresso: a modular and open-source software project for quantum
  simulations of materials},}\ }\href {http://www.quantum-espresso.org}
  {\bibfield  {journal} {\bibinfo  {journal} {J. Phys.:Condens. Matter}\
  }\textbf {\bibinfo {volume} {21}},\ \bibinfo {pages} {395502} (\bibinfo
  {year} {2009})}\BibitemShut {NoStop}%
\bibitem [{\citenamefont {Mostofi}\ \emph {et~al.}(2008)\citenamefont
  {Mostofi}, \citenamefont {Yates}, \citenamefont {Lee}, \citenamefont {Souza},
  \citenamefont {Vanderbilt},\ and\ \citenamefont {Marzari}}]{Mostofi2008}%
  \BibitemOpen
  \bibfield  {author} {\bibinfo {author} {\bibfnamefont {Arash~A.}\
  \bibnamefont {Mostofi}}, \bibinfo {author} {\bibfnamefont {Jonathan~R.}\
  \bibnamefont {Yates}}, \bibinfo {author} {\bibfnamefont {Young-Su}\
  \bibnamefont {Lee}}, \bibinfo {author} {\bibfnamefont {Ivo}\ \bibnamefont
  {Souza}}, \bibinfo {author} {\bibfnamefont {David}\ \bibnamefont
  {Vanderbilt}}, \ and\ \bibinfo {author} {\bibfnamefont {Nicola}\ \bibnamefont
  {Marzari}},\ }\bibfield  {title} {\enquote {\bibinfo {title} {wannier90: A
  tool for obtaining maximally-localised wannier functions},}\ }\href {\doibase
  10.1016/j.cpc.2007.11.016} {\bibfield  {journal} {\bibinfo  {journal}
  {Comput. Phys. Commun.}\ }\textbf {\bibinfo {volume} {178}},\ \bibinfo
  {pages} {685 -- 699} (\bibinfo {year} {2008})}\BibitemShut {NoStop}%
\end{thebibliography}%
\end{document}